\documentclass[11pt]{article}

\usepackage{verbatim}
\usepackage{amsmath}
\usepackage{amsfonts}
\usepackage{amssymb}
\usepackage{bbm}
\usepackage{latexsym}
\usepackage{graphics}
\usepackage{lscape} 
\usepackage{epsfig}
\usepackage{color}

\usepackage{cite}

\definecolor{redx}{rgb}{1,0.,0.}

\renewcommand{\d}{\mathrm{d}}

\newcommand{\Ra}{\Rightarrow}

\DeclareMathSymbol{\mg}{\mathrel}{symbols}{"1D}

\newcommand{\ga}{\alpha}
\newcommand{\gb}{\beta}
\renewcommand{\gg}{\gamma}
\newcommand{\gd}{\delta}
\renewcommand{\ge}{\epsilon}

\newcommand{\gf}{\phi}

\newcommand{\gk}{\kappa}
\newcommand{\gl}{\lambda}
\newcommand{\gr}{\rho}
\newcommand{\gvr}{\varrho}
\newcommand{\gth}{\theta}

\newcommand{\gs}{\sigma}

\newcommand{\go}{\omega}

\newcommand{\gp}{\pi}
\newcommand{\gps}{\psi}

\newcommand{\gch}{\chi}
\newcommand{\gG}{\Gamma}

\newcommand{\gL}{\Lambda}

\newcommand{\gO}{\Omega}

\newcommand{\cA}{{\cal A}}

\newcommand{\cF}{{\cal F}}

\newcommand{\cN}{{\cal N}}

\newcommand{\cR}{{\cal R}}

\newcommand{\tr}{\text{tr}}

\newcommand{\Id}{\text{\small 1}\hspace{-3.5pt}\text{1}}

\newcommand{\ra}{\rightarrow}

\newcommand{\der}{\partial}

\newcommand{\inv}{^{-1}}
\newcommand{\dsp}{\displaystyle}

\newcommand{\labl}[1]{\label{#1}}
\newcommand{\sfrac}[2]{{\scriptstyle \frac{#1}{#2}}}

\newcommand{\Kh}{K\"{a}hler}
\newcommand{\beq}{\begin{equation}}
\newcommand{\eeq}{\end{equation}}
\newcommand{\barr}{\begin{array}}
\newcommand{\earr}{\end{array}}
\newcommand{\equ}[1]{\begin{gather} #1 \end{gather}}

\newcommand{\tabu}[2]{\begin{tabular}{#1} #2 \end{tabular}}
\newcommand{\arry}[2]{\begin{array}{#1} #2 \end{array}}

\newcommand{\pmtrx}[1]{\begin{pmatrix} #1 \end{pmatrix}}

\newcounter{oldcounter}

\newcommand{\bF}{{\bar F}}

\newcommand{\bgf}{{\bar\phi}}

\newcommand{\bgps}{{\bar\psi}}

\newcommand{\bgch}{{\bar\chi}}
\newcommand{\Bgr}{{\boldsymbol \rho}}

\newcommand{\Intr}{\mathbb{Z}}
\newcommand{\Cplx}{\mathbb{C}}
\newcommand{\Real}{\mathbb{R}}

\newcommand{\Res}{\text{Res}}

\newcommand{\ba}[2]{\[\begin{array}{#2}\label{#1}}
\newcommand{\ea}{\end{array}\]}
\newcommand{\be}{\begin{equation}}
\newcommand{\ee}{\end{equation}}
\newcommand{\bea}{\begin{eqnarray}}
\newcommand{\eea}{\end{eqnarray}}

\newcommand{\U}[1]{\mathrm{U(#1)}}
\newcommand{\SU}[1]{\mathrm{SU(#1)}}
\newcommand{\SO}[1]{\mathrm{SO(#1)}}
\newcommand{\Sp}[1]{\mathrm{Sp(#1)}}

\newcommand{\rep}[1]{\mathbf{#1}}

\newcommand{\sm}{{\,\mbox{-}}}

\begin{document}

\thispagestyle{empty}

\begin{flushright}
HD-THEP-08-21 \\
%SIAS-CMTP-08-x \\
CPHT-RR075.0908\\
LPT-ORSAY-08-77 
\end{flushright}
\vskip 2 cm
\begin{center}
{\Large {\bf Non-Abelian bundles on heterotic non-compact K3 orbifold blowups} 
}
\\[0pt]

\bigskip
\bigskip {\large
{\bf Stefan Groot Nibbelink$^{a,}$\footnote{
{{ {\ {\ {\ E-mail: grootnib@thphys.uni-heidelberg.de}}}}}}},
{\bf  Filipe Paccetti Correia$^{b,}$\footnote{
{{ {\ {\ {\ E-mail: paccetti@fc.up.pt}}}}}}},
{\bf Michele Trapletti$^{c,}$\footnote{
{{ {\ {\ {\ E-mail: michele.trapletti@cpht.polytechnique.fr}}}}}}}
\bigskip }\\[0pt]
\vspace{0.23cm}
${}^a$ {\it 
Institut f\"ur Theoretische Physik, Universit\"at Heidelberg, 
Philosophenweg 16 und 19,  D-69120 Heidelberg, Germany 
\\[1ex]  
Shanghai Institute for Advanced Study, 
University of Science and Technology of China, 
99 Xiupu Rd, Pudong, Shanghai 201315, P.R.\ China
\\} 
\vspace{0.23cm}
${}^b$ {\it 
Centro de F\'isica do Porto, Faculdade de Ci\^encias da Universidade do Porto,\\
Rua do Campo Alegre, 687, 4169-007 Porto, Portugal
 \\} 
\vspace{0.23cm}
${}^c$ {\it 
Laboratoire de Physique Theorique, Bat. 210,
Universit\'e de Paris-Sud, F-91405 Orsay, France 
\\[1ex] 
Centre de Physique Th\'eorique, \'Ecole Polytechnique,
F-91128 Palaiseau, France 
 \\}

\bigskip
\end{center}

\subsection*{\centering Abstract}

Instantons on Eguchi--Hanson spaces provide explicit examples of
stable bundles on non--compact four dimensional
$\Cplx^2/\Intr_n$ %%%C^2/Z_n 
orbifold resolutions with non--Abelian structure groups.
With this at hand, we can consider compactifications of ten 
dimensional SO(32) supergravity (arising as the low energy limit of the 
heterotic string) on the resolved spaces in the presence of 
non--Abelian bundles. We provide explicit examples in the resolved 
$\Cplx^2/\Intr_3$ %%%C^2/Z_3
case, and give a complete classification of all possible effective six
dimensional models where the instantons are combined with Abelian gauge
fluxes in order to fulfil the local Bianchi identity constraint.
We compare these models with the corresponding 
$\Cplx^2/\Intr_3$ %%%C^2/Z_3
orbifold models, and find that
all of these gauge backgrounds can be related to configurations of
vacuum expectation values (VEV's) of twisted and sometimes untwisted
states.  
Gauge groups and spectra are identical from both the orbifold 
and the smooth bundle perspectives.

\newpage 
\setcounter{page}{1}

\section{Introduction} 
\label{sc:intro}

One of the central aims of string phenomenology is to construct models
that are close relatives of the Standard Model (SM) or of its supersymmetric
extension (MSSM). There have been many attempts in that direction, see
e.g.~\cite{Faraggi:1989ka,Blumenhagen:2001te,Cvetic:2001nr,Dijkstra:2004cc,Dijkstra:2004ym}, in this work we mainly focus on heterotic orbifold and 
Calabi--Yau constructions.

Orbifold compactification of the heterotic 
string~\cite{Dixon:1985jw,Dixon:1986jc,Ibanez:1987sn} 
has been one of the most successful approaches to string 
phenomenology. One of its main advantages is that strings on orbifolds
define exact CFTs and are therefore fully calculable. Many MSSM--like
models have been  
constructed~\cite{Buchmuller:2005jr,Buchmuller:2006ik,Lebedev:2006kn} 
following the route of building six dimensional intermediate 
``orbifold GUTs''~\cite{Altarelli:2001qj} from string
compactifications~\cite{Kobayashi:2004ya,Forste:2004ie,Hebecker:2004ce,Buchmuller:2005jr,Kim:2006hw}. 
But this approach has the severe limitation that away from the orbifold 
point in moduli space one quickly looses control over the resulting 
effective theory. Moving away from the orbifold point is described by 
giving vacuum expectation values (VEV's) to some twisted states, which
only makes sense when  these vevs are sufficiently small, hence one
does not have access to the full moduli space.

A generic point in the moduli space can only be described 
by giving the corresponding Calabi--Yau with a stable 
gauge bundle that it can support. This brings us to the second successful
approach to obtain the MSSM from the heterotic string as
a compactification  on elliptically fibered Calabi--Yau manifolds with
stable bundles~\cite{Donaldson:1985,Uhlenbeck:1986} on
them~\cite{Braun:2005ux,Braun:2005bw,Blumenhagen:2005zg,Bouchard:2005ag,Blumenhagen:2006ux}.  
These two procedures are very different,  hence it is very difficult
to decide whether they are closely related and give rise to the
identical models. This might well be often the case because orbifolds
are typically considered as singular limits of smooth Calabi--Yau
spaces. It is this very interesting question, how these two approaches
can be related to each other, that provides part of the inspiration
for our work.

In recent publications we have made first attempts to 
understand the relation between heterotic string orbifold 
constructions and smooth Calabi--Yau manifolds with gauge 
bundles (see \cite{Honecker:2006qz} for earlier work). To this end we have constructed explicit blowups of 
$\Cplx^n/\Intr_n$ orbifolds with Abelian gauge backgrounds satisfying
the Hermitean Yang--Mills equations. We have shown that their gauge
group and massless spectra precisely correspond to heterotic models
built on these orbifolds~\cite{Nibbelink:2007rd} (see
also~\cite{Ganor:2002ae}). Building on these 
results, we investigated the issue of multiple anomalous U(1)'s in 
blowup~\cite{GrootNibbelink:2007ew}, and how these results can be  
extended to the study of compact orbifold blowups~\cite{Nibbelink:2008tv}.  
However, generically it is not easy to obtain explicit resolutions, 
but luckily techniques of toric geometry can be employed to resolve 
many much more complicated orbifold 
singularities~\cite{Erler:1992ki,Aspinwall:1994ev}
and can even be lifted to describe the geometry of 
compact orbifold resolutions~\cite{Lust:2006zh}. To be able to also 
study the relation between heterotic strings on such generic 
orbifolds and their toric resolutions, we constructed line bundles on
them that characterize Abelian  gauge
backgrounds~\cite{Nibbelink:2007pn}. 
For essentially all the heterotic orbifold models
we  considered, we were able to find corresponding line bundle 
models, that have matching unbroken gauge groups and 
spectra (some exceptions are heterotic orbifolds without first twisted
states, where no blow up is possible.) 
These analyses show that non--compact orbifold models 
with a single twisted field taking a non--vanishing VEV along 
a supersymmetric, i.e.\ F-- and D--flat direction, that  
generates the blowup, can be matched with line bundle models 
with Abelian structure groups built on their toric resolutions.

However, line bundles only define a very small subclass of 
possible stable bundles on orbifold resolutions: 
There exist many other stable bundles that correspond to 
non--Abelian gauge backgrounds. This is also clear from the 
heterotic orbifold model perspective: Only a single of their twisted 
states takes a non--vanishing VEV to generate one of the line 
bundle models on the resolution. Clearly, there are other F-- and 
D--flat directions in which multiple twisted and untwisted states 
take non--zero VEV's simultaneously. Therefore, a more complete 
understanding of the relation between orbifold models with 
VEV's switched on and non--Abelian bundle models is 
required.

In this work we take a first step in this direction by studying this 
issue for compactifications on non-compact K3 spaces 
preserving six dimensional $N=1$ supersymmetry. 
We consider Eguchi--Hanson 
resolutions~\cite{Eguchi:1978xp,Gibbons:1979zt,Eguchi:1980jx} 
of the non--compact orbifolds $\Cplx^2/\Intr_n$, because, not 
only are these spaces known explicitly, but also a basis of all
Abelian gauge configurations have been built on them.  
In addition even a large class of non--Abelian gauge backgrounds  
have been constructed in the past~\cite{Douglas:1996sw,Bianchi:1996zj}. 
After we have reviewed the explicit constructions and discussed 
how these results can be described using a language inspired by 
toric geometry, we systematically classify all the possible resolutions 
with Abelian and non--Abelian backgrounds combined embedded in SO(32),
that  fulfill the local integrated Bianchi identity. (We focus here
mainly for simplicity only on the ten dimensional $N=1$ SO(32)
heterotic supergravity, the E$_8\times$E$_8$ can be treated
similarly.) For each of these  bundle models we are able to give the
corresponding configuration of VEV's of twisted and untwisted states
the heterotic SO(32) theory, that result in the same gauge group and
six dimensional chiral spectrum. In this sense the present paper can
be seen as the extension of the work~\cite{Honecker:2006qz} where this
matching was established for line bundles only. 
For concreteness we perform most of this study for the resolution 
of the orbifold $\Cplx^2/\Intr_3$; we are confident that our results
can be generalized to other $\Cplx^2/\Intr_n$ blowups as well.

\section{Eguchi--Hanson  $\boldsymbol{\Cplx^2/\Intr_N}$ resolutions}

In this section we give an explicit description of the resolution of
$\Cplx^2/\Intr_N$ singularities using Eguchi--Hanson spaces. After
describing the geometry we first consider Abelian gauge backgrounds on
these spaces, and then we turn to non--Abelian configurations
realized as instantons. This subsection has been based to a large
extend on~\cite{Bianchi:1996zj} (see also~\cite{ConradThesis}).

\subsection{Geometry}

The starting point of the description of Eguchi--Hanson
spaces~\cite{Eguchi:1978xp,Eguchi:1978gw,Eguchi:1980jx} in four
Euclidean dimensions is the line element  
\equ{
\d s^2 ~=~ V\inv \big( \d x_4 + \vec\go \cdot \d \vec{x} \big)^2 
\,+\, V\, d \vec{x}^2~, 
}
or equivalently the vielbein one--forms:
\equ{
\vec{e} ~=~ V^{\frac 12} \d \vec{x}~, 
\qquad 
e_4 ~=~ V^{\sm \frac 12} 
\big(\d x_4 + \vec{\go} \cdot \d \vec{x} \big)~. 
}
Here we use the three dimensional vector notation 
$\vec x^T = (x_1, x_2, x_3)\in \Real^3$, and make use of the standard
vector inner and outer products. Instead, $x_4$ has compact range, 
that will be determined below. $V$ and $\vec{\go}$ are scalar and 
vector functions of $\vec{x}$ only; we denote
derivative w.r.t.\ $x_i$, $i=1,2,3$ as $V_{,i}$, etc. The spin--connection
one--form is defined via the Maurer--Cartan structure equations   
\equ{
\d\, e_A \,+\, \gO_{AB}\, e_B ~=~ 0~, 
\qquad 
\gO_{AB} ~=~ -\, \gO_{BA}~, 
}
where $A=1,2,3,4$. A short computation shows that the independent
components read: 
\equ{
\arry{l}{ \dsp 
\gO_{4i} ~=~ \sfrac 12\, V^{\sm \frac 32}
\Big\{- V_{,i} \, e_4 - (\go_{i,j} - \go_{j,i}) e_j \Big\}~, 
\\[2ex] \dsp 
\gO_{ij} ~=~ \sfrac 12\, V^{\sm \frac 32}  
\Big\{ V_{,j} e_i - V_{,i} e_j + (\go_{i,j} - \go_{j,i}) e_4  \Big\}~.
}
\label{spinconnection}
}
The curvature two--form in turn is obtained via the conventional
expression
\equ{
R_{AB} ~=~ \d \gO_{AB} + \gO_{AC} \gO_{CB}~, 
}
The defining property of an Eguchi--Hanson space is that it has a
self--dual curvature two--form
\equ{
R_{AB} ~=~  -\frac 12\, \ge_{ABCD}\, R_{CD} ~=~ *R_{AB}~. 
}
Here $\ge_{ABCD}$ denotes the four dimensional epsilon tensor, with
$\ge_{1234}=1$. The Hodge $*$--operation acts as  
\equ{
*(e_Ae_B)   ~=~ -\frac 12\, \ge_{ABCD}\, e_Ce_D
~,
\qquad 
*^2 = \Id~, 
\labl{hodge}
}
i.e.\ 
$*(e_4\,e_i)   ~=~ \frac 12\, \ge_{ijk}\, e_j \,e_k$, given the relation
$\ge_{ijk}=\ge_{ijk4}$ between the three and the four
dimensional epsilon tensor.

A self--dual curvature is obtained automatically if the
spin--connection one--form itself is self--dual, this is
guaranteed if 
\equ{
V_{,i} ~=~ -\, \ge_{ijk}\, \go_{j,k}~
\quad \Ra \quad 
V_{,ii} ~=~ 0~. 
\labl{ASDcondition}
}
This means that $V$ is an harmonic function of $\vec{x}$. The precise
expression for this harmonic function distinguishes between
Eguchi--Hanson spaces and Kaluza--Klein monopoles: For the former the
harmonic function takes the form 
\equ{
V(\vec x) ~=~ \sum_{r=1}^N \frac{R/2}{|\vec x - \vec x_r|}~, 
\labl{pot}
}
where the points $\vec x_r$ denote the $N$ centers of the Eguchi--Hanson
space, and $R$ sets the scale of the geometry.  
(Kaluza--Klein monopoles have a similar expansion but with
an additional non--vanishing constant added.)

At the centers the function $V$ has singularities, but this does not
necessarily imply that the geometry is singular. To see this we zoom
in on one of the centers, which can be assumed to be located at the
origin, so that we can ignore the other centers, i.e.\ 
$V \ra R/(2 \gvr)$ with $\gvr = |\vec x|$. Using spherical coordinates,
\equ{
x_1=\rho\sin\theta\sin\,\phi \ ,\quad x_2=\rho\sin\theta\cos\,\phi \ ,\quad x_3=\rho\cos\theta \ ,
}
the line element for a single center can be written as 
\equ{
\d s^2\Big|_{\text{single}} ~=~ 
V\inv \big( \d x_4 + \sfrac 12 R(\cos \gth - 1) \d \gf  \big)^2
+ V \big( \d \gvr^2 + \gvr^2\, \d \gth^2 + \gvr^2 \sin^2 \gth\, \d \gf^2 \big)~,
}
which means that we have chosen a gauge in which 
\equ{
\vec\go^T ~=~ \frac{R}{2\gvr} \frac 1{\gvr+ x_3} 
\pmtrx{ x_2, -x_1, 0}~. 
}
By introducing the complex coordinates
\equ{
z_1 ~=~ \sqrt {2R}\, \gvr^{\frac 12}\, \cos(\sfrac 12 \gth) e^{ix_4/R}~, 
\qquad 
z_2 ~=~ \sqrt {2R}\, \gvr^{\frac 12}\, \sin(\sfrac 12 \gth) e^{i(\gf -x_4/R)}~, 
}
one sees that the Eguchi--Hanson space with a single center is flat
\equ{
\d s^2\Big|_{\text{single}} ~=~ 
\big| \d z_1 \big|^2 \,+\, \big| \d z_2 \big|^2~,
}
everywhere except possibly at the origin. In order that the space is
flat there as well, no deficit angle should be present, this implies
that 
\equ{
x_4 ~\sim~ x_4 \,+\, 2\gp\, R 
\labl{periodx4}
}
is periodic with a period of $2\gp\, R$. Therefore, if we want that the 
Eguchi--Hanson space has no singularities, all centers have the same 
radius $R$, as given in~\eqref{pot}.

If $n$ of the $N$ center of an Eguchi--Hanson space come close
together a $\Intr_n$ orbifold singularity arises. This can be easily
seen by reviewing the above argument when $n$ centers are on top of
each other: Indeed, the metric for this case is obtained by replacing
$R$ by $n R$. Therefore, this substitution can be made in all of the
consequent results, in particular the complex coordinates now become 
\equ{
z_1 ~=~ \sqrt {2nR}\, \gvr^{\frac 12}\, \cos(\sfrac 12 \gth) e^{ix_4/(nR)}~, 
\qquad 
z_2 ~=~ \sqrt {2nR}\, \gvr^{\frac 12}\, \sin(\sfrac 12 \gth) e^{i(\gf
  -x_4/(nR))}~,  
}
except in the periodicity~\eqref{periodx4} of $x_4$. Now, since
the Eguchi--Hanson space is non--singular when all centers are away from
each other, and this fixes~\eqref{periodx4},  when $n$ centers are on top of each
other the periodicity of $x_4$ leads to the following $\Cplx^2/\Intr_n$ orbifold
identification 
\equ{
\big( z_1, z_2 \big) ~\ra~ 
\big( z_1', z_2' \big) ~=~  
\big( e^{2\gp i/n} z_1, e^{\sm 2\gp i/n}z_2 \big)~. 
}

The complex structure that we have introduced above for the
Eguchi--Hanson space, with one or multiple centers on top of each other, 
is not unique. In fact any Eguchi--Hanson space can be equipped with
three complex structures, or a hyper--\Kh\ structure. The three \Kh\
forms,  
\equ{
J_i ~=~ 
\frac 1{\sqrt 2}\, \Big( e_4\, e_i \,-\, \sfrac 12 \ge_{ijk}\, e_j\, e_k \Big)
~=~ \frac1{\sqrt 2}\,\big(1 \,-\, *\big) e_4\, e_i~, 
}
of the hyper--\Kh\ structure are anti--self--dual, and define a Clifford
algebra 
\equ{
*\, J_i ~=~ -\, J_i~, 
\qquad 
\big\{ J_i, J_j \big\} ~=~ 2 \gd_{ij}\, \text{Vol}~, 
}
where $\text{Vol} = e_1e_2e_3e_4$ is the volume form of the
Eguchi--Hanson space.

\subsection{Abelian gauge backgrounds}
\labl{sc:AbelianEx}

An important aspect is that an Eguchi--Hanson space supports regular
Abelian gauge fluxes $\cF_r = \d \cA_r$, taken to be anti--Hermitean,
that satisfy the Hermitean--Yang--Mills equations 
\equ{
\cF_r \, J_i ~=~ 0~, 
\label{HYMequ}
}
for $i=1,2,3$ on a hyper--\Kh\ manifold. As becomes clear below, these
field strengths are labeled by $r$, the center of the Eguchi--Hanson
space. Because $J_i$ are anti--self--dual, these conditions are
identically satisfied if $\cF_r$ are self--dual, i.e.\ can be written
as  
\equ{
\cF_r ~=~ i\, F_{r\,i}\, 
\Big( e_4\, e_i \,+\, \sfrac 12 \ge_{ijk} e_j\, e_k \Big)~,  
}
for real functions $F_{r\, i}$ of $\vec x$. The closure of the field
strength of an Abelian gauge field, $\d \cF_r=0$,  implies that 
$F_{r\,i}=F_{r,i}$ for some scalar functions $F_r$. The other
components of the closure relations require these functions fulfill
the equation 
\equ{
\big[ V F_r \big]_{,ii} ~=~
V\, F_{r,ii} \,+\, 2\, V_{,i} \, F_{r,i} ~=~  0~. 
}
The first equality is obtained by using that $V$ is harmonic. Hence we
conclude that $VF_r$ is harmonic as well, and hence can be expanded in
terms of harmonic functions $1/|\vec x - \vec y|$ with constant 
$\vec y$, hence we have 
\( 
F_r(\vec x) = 1/(V(\vec x)\,  |\vec x - \vec y|). 
\) 
This means that unless $\vec y$ equals one of the positions of the
centers of the Eguchi--Hanson space, the gauge background is
singular. Therefore, we associate to each center $\vec x_r$ a gauge
background 
\equ{
\cF_r ~=~ \frac i{R}\, 
\Big( \frac {V_r}{V} \Big)_{,i} \, 
\Big( e_4\, e_i \,+\, \sfrac 12 \ge_{ijk} e_j\, e_k \Big)~, 
\quad \text{with} \quad 
V_r(\vec x) ~=~ \frac {R/2}{|\vec x - \vec x_r|}~. 
}
This field strength is obtained from the gauge connection given by 
\equ{
\cA_r ~=~ - \frac i{R}\, V^{-\frac 12}
\big[ V_r\,e_4 \,-\, \vec \go_r \cdot \vec e \big]~, 
}
where $\vec \go_r$ is defined from $V_r$ via the
equation~\eqref{ASDcondition}. The normalization of the gauge  
connections $\cA_r$ above has been chosen such that the corresponding
gauge field strengths $\cF_r$ define an orthonormal basis of
self--dual two forms~\cite{Sen:1997js,Ruback:1986ag}
\equ{
\int \frac{\cF_r \cF_s}{(2\gp)^2} ~=~ -\gd_{rs}~, 
\labl{innertwoforms}
}
where the integral is performed over the whole Eguchi--Hanson space.

Because $V = \sum_r V_r$, it follows that $\sum_r \cF_r =0$, i.e.\
only $N-1$ of these $N$ gauge backgrounds are independent. A basis of
the independent gauge backgrounds can be defined by 
\equ{
\tilde{\cF}_r ~=~ \cF_{r+1} \,-\, \cF_r~, 
}
for $r= 1,\ldots, N-1$. It follows immediately
from~\eqref{innertwoforms}, that the inner products of these
two--forms $\cF_r$ gives rise to the Cartan matrix $G$ of the
$A_{N\sm1}$ algebra of $\SU{N}$:
\equ{
\int \frac {\tilde{\cF}_r \tilde{\cF}_s}{(2\gp)^2} ~=~  -G_{rs}~. 
}
Therefore the embedding of the Abelian gauge background in the
gauge group $\SO{32}$ of the heterotic theory, is encoded by 
\equ{
\cA({\Bgr}) ~=~ \Bgr^T\, \tilde{\cA}~, 
\qquad 
\gr_r ~=~ \gr_{I\, r}\, H_I~, 
}
where $\Bgr^T = (\gr_1,\ldots,\gr_{N-1})$ is an Cartan algebra valued
vector with $H_I$ the generators of the Cartan subalgebra. We often
also view $\Bgr$ as collection of $N-1$ vector $\gr_r$ with components
$\gr_r^I$.

\subsection{Non--Abelian gauge backgrounds}
\label{sc:NonEx}

Eguchi--Hanson spaces also support non--Abelian gauge backgrounds. The
tangent bundle obviously defines an example of a non--Abelian gauge
background on this space. In this section we would like to review how
a large class of non--Abelian fluxes, or instantons, can be constructed
explicitly. Such instantons are generalizations~\cite{Bianchi:1996zj} of the
't~Hooft instantons~\cite{'tHooft:1974qc} on $\Real^4$. We first consider
$\SU{2}$ gauge background and then at the end of this subsection
comment how to construct gauge backgrounds with other structure
groups.

Consider a gauge connection one--form 
\equ{
\cA ~=~ i\, V^{-\frac 12}\, 
\Big[ A_4\, e_4 \,+\, \vec A \cdot \vec e \Big]~, 
}
which takes values in the $\SU{2}$ algebra generated by the
Pauli--matrices $\gs_i$. In order that the corresponding  non--Abelian
gauge field strength 
\(
\cF = \d \cA + \cA^2
\)
satisfies the Hermitean--Yang--Mills equations~\eqref{HYMequ}, it has to
be self--dual as the Abelian gauge backgrounds discussed in the
previous subsection. This implies that the matrix-valued one-forms $A_4$ and
$\vec A$ satisfy 
\equ{
- A_{4,i} \,+\,  i[A_4,A_i] ~=~ \ge_{ijk} \, 
\Big( - A_{j,k} \,+\, \sfrac i2\,[A_j,A_k] \Big)~. 
}
To solve this equation we make the ansatz for the potential one--forms
\equ{
A_4 ~=~ P_i \, \gs_i~, 
\qquad 
A_i ~=~ - \ge_{ijk}\, P_j \, \gs_k~, 
}
where $P_i$ are scalar functions to be
determined. Substituting this ansatz into the equation above, leads to
two independent relations
\equ{
P_{i,j} \,-\, P_{j,i} ~=~ 0~, 
\qquad 
P_{i,i} - 2 (P_i)^2 ~=~ 0~. 
}
The first identity implies that $P_i = P_{,i}$ of a single scalar
function $P$; the second equation implies that this can be expressed as 
\equ{
P(\vec x) ~=~ - \sfrac 12 \, \ln H(\vec x)~, 
}
where $H$ is again an harmonic function. The centers of this harmonic
function have to coincide with some of the centers of the
Eguchi--Hanson space, otherwise the background is a configuration that
does not have finite action, i.e.\ is singular. We will often say that
the harmonic function $H$ and therefore the corresponding instanton
are supported at some of the centers of the Eguchi--Hanson space. 
To summarize, the gauge background becomes
\equ{
\cA ~=~ - i\, V^{-\frac 12}\, 
\Big\{ \frac {H_{,k}}H\, e_4 \,+\, \ge_{ijk} \frac {H_{,i}}H\,e_j \Big\}
\, \sfrac 12 \, \gs_k 
~=~ 
- V^{-\frac 12}\, \frac{H_{,A}}H\, e_B \,  \sfrac 12 \, \gg_{AB}^+~, 
\label{nonAbelian}
}
where after the second equal sign we have used the four component
spinor notation of $\SO{4}$ to emphasize that the non--Abelian bundle only
affects the positive chirality sector. (For our conventions concerning
spinor representation properties see Appendix~\ref{sc:clifford}.)
Its field strength reads
\equ{
\cF ~=~ \sfrac i2\, V\inv 
\Big\{ 
\Big( \frac{H_{,ij}}{H}
\,-\, 2\, \frac{H_{,i}}{H} \frac{H_{,j}}{H}
\,-\, \frac{V_{,i}}{V} \frac{H_{,j}}{H}
  \Big)\, \gs_j
\,+\, \frac{H_{,m}{}^2}{H^2}\, \gs_i
\Big\} \, 
\big( e_4 e_i \,+\, \sfrac 12\, \ge_{ikl}\, e_k e_l \big)~.
}

As a first important example of a non--Abelian gauge background, we
consider the standard embedding in which the gauge connection is
determined by the spin--connection
\equ{ 
\cA_{SE} ~=~ 
\Big(  
\gO_{4k} \,+\, \sfrac 12\, \gO_{ij}\, \ge_{ijk}
\Big) \, \sfrac i2 \, \gs_k~. 
}
By comparing the expressions for $\gO_{4i}$ and $\gO_{ij}$ given
in~\eqref{spinconnection} and the generic non--Abelian gauge
background~\eqref{nonAbelian}, we infer that for the standard
embedding we have $H(\vec x) = V(\vec x)$. Therefore the standard
embedding is a non--Abelian gauge background that has support at all
centers of the underlying Eguchi--Hanson space. Other non--Abelian
gauge backgrounds are not supported at all Eguchi--Hanson centers.

The non--Abelian gauge backgrounds above are classified by their instanton
numbers
\equ{
\int c_2(\cF) ~=~ \int \, \sfrac 12\, \tr \Big(\frac {\cF}{2\gp i} \Big)^2~, 
\label{instantonNr}
}
obtained as integrals over the second Chern class, for this see
e.g.~\cite{Nakahara:1990th} (moreover, $c_1(\cF) =0$). The instanton number
is related to the number $p$ of Eguchi--Hanson centers where a
non--Abelian gauge flux has support.  To determine this relation, we
make the following observations: Away from the centers, the gauge
configuration is pure gauge, hence the field strength vanishes
there. Therefore, the only contributions to the instanton number come from
the centers of the non--Abelian background and the asymptotic for  
$\vec x \ra \infty$. To compute the contribution from the centers, we
consider a small ball $B_{\vec x_I}$ surrounding the center $\vec x_I$,
and we use Stoke's theorem 
\equ{
\int_{B_{\vec x_I}} c_2(\cF) ~=~ 
\int_{\der B_{\vec x_I}} \go_{CS}(\cA) ~=~ 
\frac 1{8\gp^2}
\int_{\der B_{\vec x_I}} \sfrac 13\, \tr \cA^3  ~=~ 1~.  
}
Here we used that only the second term of the Chern--Simons
three--form 
\(
\go_{CS}(\cA) = -\tr( \cF \cA - \frac 13 \cA^3)/(8\gp^2)
\)
does not vanish. This computation holds for each center separately,
when all centers are at finite distance from each other. Because this
is a topological quantity even in the limit when $p$ centers come
close together, each of them still has an instanton number $1$, hence
collectively they have instanton number $p$. The instanton number at
infinity can be computed in a similar way, but now only the leading
contributions have to be taken into account. For an
Eguchi--Hanson space with an instanton that is 
supported at $p$ of its $N$ centers this means that 
\equ{
V(\vec x) ~=~ \frac {NR}{2|\vec x|}~, 
\qquad 
H(\vec x) ~=~ \frac {pR}{2|\vec x|}~, 
}
for large $|\vec x|$. Since $H$ only appears in a logarithm, that
determines the non--Abelian gauge connection, the
pre--factor in $H$ is in fact irrelevant. Hence,  the integral over
the region $X_\gvr = \{\vec x, |\vec x| > \gvr\}$ gives, using Stoke's,
\equ{
\int_{X_r} c_2(\cF) ~=~ -\frac 1{N}~,
}
when $\gvr\ra\infty$ because the orientation is opposite w.r.t.\ that
around the centers of the instanton.  Collecting the various
contributions we conclude that the instanton number of an instanton
with support at $p$ of its $N$ centers of a Eguchi--Hanson space is
given by
\equ{
\int c_2(\cF) ~=~ p \,-\, \frac 1N~. 
\label{2ndChern}
}

The instantons discussed so far only define $\SU{2}$ gauge
configurations, instantons in other gauge representations can be
easily obtained from these. A complete and general investigation of
instantons on Eguchi--Hanson spaces involves a combined
ADHM~\cite{Atiyah:1978ri,Dorey:2002ik} and
Kronheimer--Nakajima~\cite{Kronheimer:1989zs,Kronheimer:1990,Nakajima:1990}
construction,  for a comprehensive review see
e.g.~\cite{Bianchi:1996zj,Fucito:2004ry}. We 
make use of an easier but less general approach~\cite{Wilczek:1976uy}
(reviewed in~\cite{Vandoren:2008xg}) in which the spin--$\frac 12$ generators
$\frac  12 \gs_i$ of $\SU{2}$ are replaced by generators $T_i$ in a
generic representation of $\SU{2}$ in the expressions for the gauge
background~\eqref{nonAbelian}. In particular the instanton
number~\eqref{instantonNr} in that representation is obtained by
replacing $\tr\big( \sfrac 12 \gs_i \sfrac12 \gs_j \big)$ by 
$\tr\big(T_iT_j)$. An irreducible representation $\rep{R_j}$ is 
labeled by the spin quantum number $j = 0, \frac 12, 1, \frac 32,$
etc.; its dimension and quadratic Casimir are given by
$\dim \rep{R}_j = 2j+1$ and $C_j = j(j+1)$ respectively. Therefore,
the instant number of representation $\rep{R}_j$ is 
\equ{
k_j ~=~ \frac 23\, C_j \, \dim \rep{R}_j ~=~ 
\frac 23 \, j(j+1)(2j+1) 
}
times larger than that in the fundamental spin--$\frac 12$
representation. If we embed a spin--$j$ representation in $\SU{M}$
with $M\geq 2J+1$, a SU(2$j$+1) subgroup is filled up, hence the
subgroup SU($M$-2$j$-1) remains unbroken.

For the embedding of instanton configurations in $\SO{32}$ groups,
which is of main interest in this paper on heterotic $\SO{32}$ blowup
models, it is important to realize that 
\(
\SO{4} = \SU{2}_+\times \SU{2}_-
\)  
on the level of the algebra, where the $\pm$ on the $\SU{2}$s refer to
the chiralities of the spinor representations. Explicit
representations of the $\SU{2}_\pm$ are $\gg_{AB}^\pm$ defined in
Appendix~\ref{sc:clifford}. Hence using the
spin--$\frac 12$ configuration we the symmetry breaking pattern reads 
\equ{
\SO{32} ~\ra~ \SO{28} \times \SU{2}_+   \times \SU{2}_-
~\ra~ 
\SO{28} \times \SU{2}_-~,  
}
because the gauge background has positive chirality,
see~\eqref{nonAbelian}. When we consider the embedding of a second
identical spin--$\frac 12$ instanton, the chirality forces us to
embed it in the $\SO{28}$. The surviving gauge group in this case is 
$\SO{24}\times\Sp{4}_-$. The explicit representation of the 
generators of this symplectic group is given in
Appendix~\ref{sc:clifford}. Similarly, when we have a triple or 
quadruple embedding of identical instantons, we obtain the left--over
symmetry groups $\SO{20}\times\Sp{6}_-$ and $\SO{16}\times\Sp{8}_-$,
respectively. Finally, it is possible to use the spin--1 embedding
into $\SO{32}$, because this representation is a vector representation,
it induces the symmetry breaking to $\SO{29}$.

\section{Toric $\boldsymbol{\Cplx^2/\Intr_N}$ resolutions}
\label{sc:resC2Zp}

We review the resolution Res($\Cplx^2/\Intr_N$) described using toric
geometrical terms, and give a systematic account of gauge fluxes on
such resolutions.  This section is based in part
on~\cite{Lust:2006zh,Nibbelink:2007pn}. (For a more detailed account
on toric geometry, see e.g.~\cite{Fulton,Oda,Bouchard:2007ik}.)

\subsection{Geometry}

Let $z_1, z_2$ denote the coordinates of $\Cplx^2$ associated
with those of the orbifold $\Cplx^2/\Intr_N$ before the blowup, and 
$x_1,\dots x_r$, $r=1,\ldots N-1$ the additional homogeneous
coordinates that define the toric variety 
\equ{
\text{Res}(\Cplx^2/\Intr_N) ~=~ 
\Big( \Cplx^{N+1} - \{0\} \Big) \,\Big/\, (\Cplx^*)^{N-1}~. 
}
The extra homogeneous coordinates $x_r$ are associated with the
twisted sectors $w_r = (r,N-r)/N$ of a $\Cplx^2/\Intr_N$ orbifold
theory. The local coordinates constructed from the homogeneous ones
\equ{
Z_1 ~=~ z_1   \prod_{r=1}^{N-1} x_r^{(N-r)/N}~, 
\qquad
Z_2 ~=~ z_2 \prod_{r=1}^{N-1} x_r^{r/N}~, 
}
are invariant under the complex scalings:
\equ{
\arry{rcl}{ 
\big(z_1,x_1,x_2\big) &\sim& \big(\gl_1\,z_1,\gl_1^{-2}\,x_1,\gl_1\,
x_2\big)~,
\\ &\vdots& \\
\big(x_{N-2},x_{N-1},z_2\big) &\sim& \big(\gl_{N-1}\,x_{N-2},\gl_{N-1}^{-2}\,x_{N-1},\gl_{N-1}\,
z_2\big)~,
}
}
where $\gl_1, \ldots, \gl_{N-1} \in \Cplx^*$.

The ordinary and exceptional divisors are defined as 
$D_i = \{ z_i = 0 \}$, $i=1,2$,  and $E_r = \{ x_r = 0 \}$, 
$r = 1,\ldots, N-1$, respectively. The exceptional divisors are
compact, while the ordinary ones are not. From the fan of the toric
diagram we read off the intersections  
\equ{
E_r E_{r+1} ~=~ 1~, 
}
for $r = 0,\ldots N$, when we write $E_0 = D_2$ and $E_N = D_1$. The
self--intersections of the exceptional divisors equal
\equ{
E_r^2 ~=~ -2~, 
}
with $r=1,\ldots, N-1$. The intersections of the exceptional divisors
can be conveniently grouped together as:
\equ{
{\bf E} \, {\bf E}^T ~=~ -G~. 
\label{CartanMatrix}
} 
where $G = G(A_{N-1})$ is the Cartan matrix of $\SU{N}$ and 
${\bf E}^T = (E_1,\ldots,E_{N-1})$. The ordinary
divisors are not independent from the exceptional ones because of the
following linear equivalence relations 
\equ{
D_1 ~\sim~  - \sum_{r=1}^{N-1} \, \frac rN \, E_{r}~, 
\qquad 
D_2 ~\sim~ - \sum_{r=1}^{N-1} \, \frac{N-r}N \, E_{r}~. 
\labl{linequiv}
}

These relations are compatible with the (self--)intersections given
above, and can be used to show that the Euler number of the resolution
is given by 
\equ{
\gch(\text{Res}(\Cplx^2/\Intr_N)) ~=~ 
\int c_2(\text{Res}(\Cplx^2/\Intr_N)) ~=~ N \,-\, \frac 1N~.  
}
To obtain this one may expand to second order the total Chern class 
represented as a product over all divisors
\equ{
c(\Res(\Cplx^2/\Intr_N)) ~=~ 
(1+D_1) (1+D_2) 
\prod_{r=1}^{N-1} (1+E_r)~, 
\label{Chtot}
}
and use the intersection numbers are described above. 
If one expands the total Chern class to first order and uses the
linear equivalence relations~\eqref{linequiv}, one finds zero. This
shows that the space has vanishing first Chern class, i.e.\ a
non--compact four dimensional Calabi--Yau.

\subsection{Abelian gauge fluxes}
\labl{sc:AbelianToric}

Next we turn to describe Abelian gauge configurations on the
resolution of the $\Cplx^2/\Intr_N$ singularity. As an Abelian gauge flux
$\cF$ can be expanded in terms of the exceptional divisors, we may
write 
\equ{
\frac{\cF}{2\gp} ~=~ \Bgr^T {\bf E} ~=~ \gr_1 E_1 + \ldots + \gr_{N-1} E_{N-1}~, 
}
for some coefficients $\gr_r$ inside the vector 
$\Bgr^T = \big( \gr_1,\ldots, \gr_{N-1} \big)$. These coefficients have to
be chosen such that the gauge flux is properly quantized. This means
that the entries of the vector  
\equ{
{\bf Q} ~=~ - \int {\bf E}\, \frac{\cF}{2\gp} ~=~ G\, \Bgr~, 
\label{QuantVector}
}
are all ``charges'',  i.e.\ elements $Q_r \in \gL$, of the lattice
spanned by vectorial and spinorial weights of SO(32). Any choice of
the charges constitutes a valid gauge background 
\(
\frac{\cF}{2\gp} ~=~ {\bf Q}^T \,G\inv\, {\bf E}~, 
\)
resulting in a contribution to the Bianchi identity
\equ{
-\frac 12 \int \Big(\frac{\cF}{2\gp} \Big)^2 ~=~ \frac 12\, \Bgr^T \, G \, \Bgr
~=~ \frac 12\, {\bf Q}^T \, G\inv\, {\bf Q}~.
\label{BianchiPart}
}
On the resolution, the orbifold gauge shift vector $v$ can be computed
as the flux around one of the coordinate axes, i.e.\ integrals over
the divisors $D_i$. This identification has to hold only up to
vectors out of the lattice $\gL$, denoted by ``$\equiv$''. Because the
orientation of the orbifold action on the coordinates $z_1$ and $z_2$
is opposite, we have 
\equ{
- v ~\equiv~ \int_{D_2}  \frac{\cF}{2\gp}  ~=~ 
\gr_1 ~=~ 
\frac 1N\, \sum_{r=1}^{N-1} r \, Q_{p\sm r}~, 
\quad 
v ~\equiv~ \int_{D_1} \frac{\cF}{2\gp}  ~=~ 
\gr_{N\sm 1} ~=~ 
\frac 1N\, \sum_{r=1}^{N-1} (N- r) \, Q_{p\sm r}~. 
}
Either of these equations tells us that $v$ is properly quantized in
units of $1/N$,  and that they are compatible because 
\equ{
\gr_1 ~+~ \gr_{N-1} ~=~ \sum_{r=1}^{N-1} Q_r ~\in~ \gL~
}
equals a lattice vector in any case. Therefore any choice of charges
${\bf Q}$ defines a consistent gauge background that can be identified
with orbifold boundary conditions in the blow down limit.

To find the properly quantized $\Bgr$ is not so straightforward in
general. Since in the latter part of this paper we focus on models on
the resolution of $\Cplx^2/\Intr_3$ we remind the reader of the
properly quantized bases found previously~\cite{Nibbelink:2007pn}
\equ{
\frac {\cF_V}{2\gp} ~=~ (V^I_1\, D_1 \,+\, V^I_2\, D_2) \, H_I~,
\labl{AbelianBack}
}
where $V_1$ and $V_2$ are vectorial or spinoral lattice vectors. 
Upon converting the $D$'s to the $E$'s and using the linear
equivalence relations, we see that this means that 
\equ{
\gr_1 ~=~ - \frac 13 \big( V_1 \,+\, 2\, V_2 \big)~, 
~~ 
\gr_2 ~=~ - \frac 13 \big( 2\, V_1 \,+\, V_2 \big)~. 
}
The contribution to the Bianchi identity then reads
\equ{
- \frac 12 \int \tr \Big( \frac{\cF_V}{2\gp} \Big)^2 ~=~ 
\frac 13 \, \big( V_1^2 \,+\, V_2^2 \,+\, V_1\cdot V_2 \big)~.  
}

\subsection{Relation with explicit construction of (non--)Abelian gauge fluxes} 
\labl{sc:NonToric}

\begin{figure}
\begin{center}
\raisebox{0cm}{\scalebox{.6}{\mbox{\input{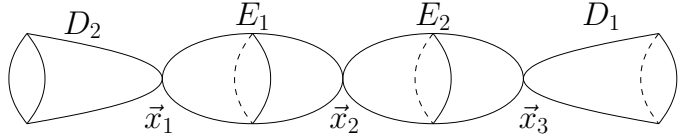}}}}
\end{center}
\caption{
Schematic picture of the compact and non--compact curves within the  
resolution of $\Cplx^2/\Intr_3$ corresponding to the exceptional
divisors $E_r$ and the ordinary divisors $D_i$, respectively. 
\label{fg:curves}}
\end{figure}

In the previous section we have discussed explicit solutions of the
non--compact Calabi--Yau condition and presented explicit
constructions of Abelian and non--Abelian gauge backgrounds. 
Comparing the results of the Abelian gauge fluxes and
the construction of the divisors shows, that we can make 
identifications between the exceptional and ordinary divisors and the
characteristic classes corresponding to the gauge field strength
(denoted by $[\ldots]$)
\equ{
2\gp\, E_r ~=~ [\tilde{\cF}_r] ~=~ [\cF_r \,-\, \cF_{r+1}]~, 
\qquad 
2\gp\, D_1 ~=~ [\cF_N]~. 
\qquad 
2\gp\, D_2 ~=~ -[\cF_1]~. 
}
By the Poincar\'e duality we know that the divisors also have an
interpretation as complex curves in the resolution space. For this 
we assume that all the centers $\vec x_r$, $r=1,\ldots N\sm 1$,  lie
ordered on one line. The representation of the exceptional divisors are 
two--spheres suspended at two adjacent centers~\cite{Bianchi:1996zj} 
\equ{
E_r ~=~ \Big\{
\big(\vec x, x_4\big) ~\big|~x_4 ~\in~ [0,2\gp R[~, 
~~ \vec x ~=~ \vec x_r + \gl\,\big( \vec x_{r+1}-\vec x_r\big)~,
\gl ~\in~ [0,1]~
\Big\}~. 
}
Clearly these surfaces are compact, and only nearest neighbor
divisors have non--vanishing intersection number one, as they
intersect only at a single point: the center that they both have in
common. In a similar way we can also give a representation of the 
non--compact ordinary divisors
\equ{
\arry{c}{\dsp  
D_1 ~=~ \Big\{
\big(\vec x, x_4\big) ~\big|~x_4 ~\in~ [0,2\gp R[~, 
~~ \vec x ~=~ \vec x_N \,+\, \gl\,\vec e_3~,
\gl ~\geq~ 0~
\Big\}~, 
\\[1ex] \dsp 
D_2 ~=~ \Big\{
\big(\vec x, x_4\big) ~\big|~x_4 ~\in~ [0,2\gp R[~, 
~~ \vec x ~=~ \vec x_1 \,-\, \gl\,\vec e_3~,
\gl ~\geq~ 0~
\Big\}~, 
}
}
Hence, the intersections $D_1 E_{N-1} = D_2 E_1 = 1$ are consistent with
what we found before. The Abelian gauge fluxes are thus associated
with the complex curves between two centers of an Eguchi--Hanson space
A schematic picture of these curves and their intersections is sketched 
in Figure~\ref{fg:curves}.

Non--Abelian bundles on Eguchi--Hanson spaces we can describe by
similar pictures. As we have seen in Subsection~\ref{sc:NonEx},
instantons on Eguchi--Hanson spaces are supported at one or more
centers of the Eguchi--Hanson space. In particular, the standard
embedding is supported on all centers, and therefore all divisors participate
to the total Chern class~\eqref{Chtot}: Precisely because the
standard embedding instanton is supported at each of the centers we
cannot deform the curves at these points.

\begin{figure}
\begin{center}
\raisebox{0cm}{\scalebox{.6}{\mbox{\input{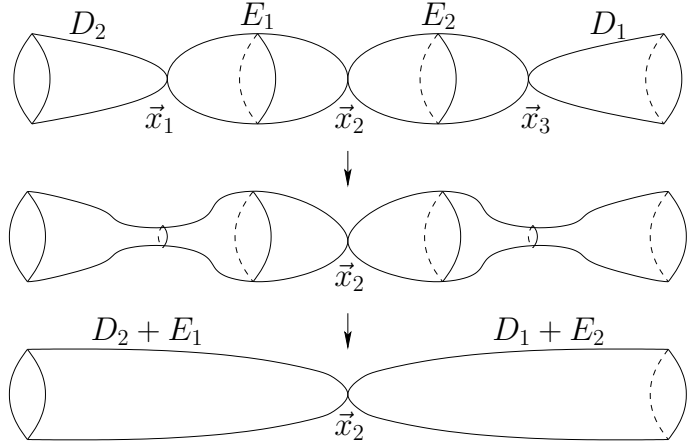}}}}
\end{center}
\caption{
The two curves $D_2, E_1$  and $D_1,E_2$ are merged to form the 
curves $D_2+E_1$ and $D_1+E_2$, respectively. 
\label{fg:mergecurves}}
\end{figure}

Instead, for the instanton $I_{\vec x_2}$ supported only at 
$\vec x_2$, we can merge the curves $D_2$ and $E_1$, because there is
no obstruction created by the instanton. The resulting curve is
denoted as $D_2+E_1$. Similarly the curves $D_1$ and $E_2$ can be 
merged to form $D_1+E_2$. This process is depicted in 
Figure~\ref{fg:mergecurves} for the $\Cplx^2/\Intr_3$ singularity
given in Figure~\ref{fg:curves}.  Therefore, as far as the
instanton supported only at $\vec x_2$ is concerned, there are only
two divisors $D_2+E_1$ and $D_1+E_2$ relevant, and consequently
its total Chern class reads
\equ{
c(I_{\vec x_2}) ~=~ (1+D_2+E_1)(1+D_1+E_2)~. 
}
Because this describes an SU(2) (non--Abelian) flux the first Chern class
vanishes identically, as follows directly from expanding this to first
order and using the linear equivalence relations~\eqref{linequiv}. For
the second Chern class we find 
\equ{
\int c_2(I_{\vec x_2}) ~=~ 1 \,-\, \frac 1N~,
}
using the intersection numbers given above. This is consistent with
the result computed in~\eqref{2ndChern} using the explicit instanton solution
on the Eguchi--Hanson space. One can check that also for instantons
supported at multiple centers this procedure gives the correct value
$p-1/N$ for the second Chern class, and that this result only depends 
on the number $p$ of centers present in the instanton, not at their location.

\section{Blowup models on non--compact K3 orbifolds}

In the previous two sections we used both explicit constructions and
implicit toric geometry methods to describe the geometry of 
non--compact resolutions of $\Cplx^2/\Intr_N$ orbifolds, and the
Abelian and non--Abelian gauge configurations they can support. 
The purpose of this section is to show that the resulting models
can be understood as non--compact heterotic orbifold models with
certain VEV's switched on. For concreteness we restrict ourselves to 
models on the $\Cplx^2/\Intr_3$ orbifold only. The corresponding
heterotic orbifold models are listed in Table~\ref{tb:HetC2Z3}. Below
we list the possible smooth models obtained by combining the Abelian
and non--Abelian bundles constructed in the previous sections. Since
by definition all these configurations are  supersymmetric as the
gauge backgrounds were required to satisfy the Hermitian Yang-Mills
equations, we restrict ourself to the blow-ups of heterotic
$\Cplx^2/\Intr_3$ orbifold models that do not break supersymmetry, 
thus, we only consider VEV's along flat directions of the  potential. 

We stress the fact that the models we consider are non--compact, but they
are built in a way such that compact (global) orbifolds can be recovered
in the simplest possible way. 
In particular, we enforce on the local models all the conditions required in 
the global models, in this way we have that the spectrum of $T^4/\Intr_3$
models can be obtained by just trivially summing over its 9 $\Cplx^2/\Intr_3$
singularities, i.e. by multiplying by 9 the $\Cplx^2/\Intr_3$ spectra given in
Table~\ref{tb:HetC2Z3}.

order that a spectrum 

\begin{table}[t]%\footnotesize
\renewcommand{\arraystretch}{1.2}
\begin{center}
\begin{tabular}{|c||c||c|c|}
\hline
$\!\!$\#$\!\!$ & 
\tabu{c}{Gauge group \\  Shift vector} 
& 
\tabu{c}{Untwisted \\ matter} 
& 
\tabu{c}{Twisted \\ matter} 
\\\hline\hline
3a& $SO(28) \times SU(2) \times U(1)$ & 
$\frac{1}{9}\left[({\bf 28},{\bf 2})_1+1({\bf 1},{\bf 1})_2+2({\bf 1},{\bf 1})_0\right]$ & 
$({\bf 28},{\bf 2})_{-1/3}+5({\bf 1},{\bf 1})_{2/3}$ % $+18({\bf 1},{\bf 1})_{4/3}$
\\
& $\frac{1}{3}(1^2,0^{14})$ & &  $+2({\bf 1},{\bf 1})_{4/3}$
\\\hline
3b & $SO(22) \times SU(5) \times U(1)$ & 
$\frac{1}{9}\left[({\bf 22},{\bf 5})_1+({\bf 1},{\bf 10})_2+2({\bf 1},{\bf 1})_0\right]$ 
& $ ({\bf 22},{\bf 1})_{5/3} + ({\bf 1},{\bf 10})_{-4/3}$ 
\\
& $\frac{1}{3}(1^4,2,0^{11})$ & &$+2({\bf 1},{\bf 5})_{-2/3}$ 
\\\hline
3c & $SO(16) \times SU(8) \times U(1)$ & 
$\frac{1}{9}\left[({\bf 16},{\bf 8})_1+({\bf 1},{\bf 28})_2+2({\bf 1},{\bf 1})_0\right]$
& $({\bf 1},{\bf 28})_{-2/3}+2({\bf 1},{\bf 1})_{8/3}$
\\
& $\frac{1}{3}(1^8,0^8)$ & & 
\\\hline
3d & $SO(10) \times SU(11) \times U(1)$ & 
$\frac{1}{9}\left[({\bf 10},{\bf 11})_1+({\bf 1},{\bf 55})_2+ 2({\bf 1},{\bf 1})_0\right]$ &
$({\bf 1},{\bf 11})_{-8/3}+(\overline{\bf 16},{\bf 1})_{-11/6}$
\\
& $\frac{1}{3}(1^{10},2,0^5)$ & & 
\\\hline
3e & $SU(14) \times SU(2)^2 \times U(1)$ & 
$\frac{1}{9}\left[({\bf 14},{\bf 2},{\bf 2})_1+({\bf
  91},{\bf 1},{\bf 1})_2+2({\bf 1})_0\right]$ 
& $({\bf 1})_{14/3} + ({\bf 14},{\bf 2},{\bf 1})_{-4/3}$
\\
& $\frac{1}{3}(1^{14},0^2)$ & &  $+2({\bf 1},{\bf 1},{\bf 2})_{-7/3}$ 
\\\hline
\end{tabular}
\end{center}
\caption{SO(32) heterotic orbifold spectra on
$\Cplx^2/\Intr_3$, see e.g.~\cite{Aldazabal:1997wi,Ibanez:1997td,Honecker:2006qz}.}
\label{tb:HetC2Z3}
\renewcommand{\arraystretch}{1}
\end{table}

\subsection{Abelian and non-abelian bundles on the resolved
$\Cplx^2/\Intr_3$ singularity}

We consider a smooth resolution of the $\Cplx^2/\Intr_3$ orbifold. The
Abelian and non--Abelian gauge configurations were discussed at length
in subsections~\ref{sc:AbelianEx}, \ref{sc:NonEx}
and~\ref{sc:AbelianToric}, \ref{sc:NonToric} as explicit and toric
geometrical constructions, respectively. Their collective
characterization can be summarized as follows: The vectors $V_1$ and
$V_2$ define the embedding of the two line bundles present in the
resolution, see~\eqref{AbelianBack}. 
The number $n_{p}^{1/2}$ counts the number of $\SU{2}$
bundles embedded in $\SO{32}$ supported at $p$ centers on the 
Eguchi--Hanson resolution space ($p=1,\,2,\,3$ because we treat
the $\Intr_3$ case). Finally $n_{p}^{1}$ is defined similar to
$n_{p}^{1/2}$, but the spin--one representation of $\SU{2}$ is used
instead.

To obtain non--compact resolution models, for which we are readily able 
to compute spectra, we enforce the local Bianchi identity 
\equ{ 
\frac 13\, \big( V_1^2\,+\, V_2^2 \,+\, V_1\cdot V_2 \big)
~+~ \frac 43\, n_1^1 
~+~ \frac 23\, n_{1}^{1/2}  \,+\, \frac 53\, n_{2}^{1/2}\,+\,\frac 83\, n_{3}^{1/2}
~=~ K~, 
\quad K~=~  \frac 83~. 
\labl{localBI}
}
Other possible bundles could be present if we did not require the local
Bianchi identity to be fulfilled, i.e.\ when $K \neq 8/3$. (The 
Bianchi--identity here is given in the normalization appropriate for 
SU--groups. The spin--1 embedding of the instanton is defined in 
SO(3), hence the factor in front of $n_1^1$ is $4/3$ rather than
$4*2/3$. The relevant integrals are summarized in
Table~\ref{tb:Multiplicities}.)  Since the contributions involving
$V_1$ and $V_2$ always give a non--negative contribution, 
the instanton numbers of the (non--)Abelian configurations 
\(\smash{
n = (n_1^1,n_{1}^{1/2}, n_{2}^{1/2},n_{3}^{1/2})
}
\)
satisfy $4 n_1^1+2 n_{1}^{1/2}  + 5 n_{2}^{1/2}+8 n_{3}^{1/2}\le 8$. 
The possible configurations are listed in Table~\ref{classnonab1}. If
this sum equals eight, only non--Abelian bundles are involved; if it
vanishes only line bundles are employed; otherwise a mixture of both
types is required. The identification of the line bundle vectors with
the $\Intr_3$ orbifold shift imposes that $V_1/3 = - V_2/3$ up to the
addition of lattice vectors~\cite{Nibbelink:2007pn}. In fact in most cases no
lattice vectors are needed, thus, in general we have
$V_1=-V_2=(0^{16-m_1-m_2},1^{m_1},2^{m_2})$ with $m_1+m_2 < 16$, 
or $V_1=-V_2=\frac{1}{2}(1^{m_1},3^{m_2})$ and $m_1+m_2=16$. 
One additional constraint is that the line bundle vectors are properly
quantized such that the Freed--Witten
anomaly~\cite{Witten:1985xe,Freed:1986zx} does 
not arise: The first Chern class of the bundle, i.e.\ the sum of the
entries of the line bundle vectors, needs to be even. From
equation~\eqref{localBI} it follows immediately, that if
$\smash{n_2^{1/2}} =1$ (or odd in general) then 
this condition is violated.  
Finding the relevant $m_1$ and $m_2$ is straightforward and the
results are listed in the second column of Table~\ref{classnonab1}.

Given the topological characterization of the Abelian and non--Abelian
bundles, the gauge symmetry breaking they induce can be investigated. 
When the Abelian and non--Abelian gauge fluxes are embedded in
different parts of $\SO{32}$, the resulting unbroken gauge group is
the intersection of the groups that are unbroken by either flux. The
other possibility is that the Abelian and the non--Abelian gauge
backgrounds share some Cartan generators; they ``overlap''. Because
the two types of fluxes commute with each other, if the $\SU{2}$
instanton has a Cartan generator, say $H_1+H_2$, and non--Abelian
generators corresponding to the weights $\pm (1^2,0^{14})$, then the
Abelian flux has to be embedded in the $\SO{32}$ Cartan as $H_1-H_2$. 
This Cartan ``overlapping'' of the non--Abelian and Abelian gauge flux has
two consequences: The unbroken SO($N$) group will be larger, while
the Sp($2n$) or $\SU{2}$ group are (partially) broken. The amount of
Cartan ``overlap'' of the Abelian flux with the instanton gauge
configuration is indicated in the overbraced part of the line bundle
vector $V_1$. The resulting unbroken gauge 
group is listed in the third column of Table~\ref{classnonab1}. 
We stress that a specific overlap, treated in \cite{Honecker:2006dt},
is the case of a $U(2)$ bundle embedded in $\SO{32}$.

\begin{table}
\begin{center} 
\begin{tabular}[b]{|ccc|}
\hline&&\\[-2ex]
spin--$\frac 12$ ~~ SU(2)$_+$ & & spin--1 ~~ SO(3)
\\[1ex] 
\begin{tabular}[t]{|c|c|}
\hline&\\[-2ex]
$\rep{r}_\gk$ & $\frac 1{8\gp^2} \int \tr_{\rep{r}_\gk} (\cF')^2$ 
\\[1ex]\hline\hline &\\[-2ex]
$\rep{1}$ & $0$ 
\\[1ex]
$\rep{2}$ & $p \,-\, \frac 13 $
\\[1ex]
$\rep{3}$ & $4( p  \,-\, \frac 1{3})$
\\[1ex]\hline
\end{tabular}
& \quad & 
\begin{tabular}[t]{|c|c|}
\hline&\\[-2ex]
$\rep{r}_\gk$ & $\frac 1{8\gp^2} \int \tr_{\rep{r}_\gk} (\cF')^2$ 
\\[1ex]\hline\hline &\\[-2ex]
$\rep{1}$ & $0$ 
\\[1ex]
$\rep{3}$ & $4( p  \,-\, \frac 1{3})$
\\[1ex]\hline
\end{tabular}
\\[-2ex] && \\[1ex]\hline
\end{tabular} 
\end{center} 
\caption{
Depending on the representation under the SU(2) or SO(3)  group
characterizing the embedding of the instantons the multiplicities of
zero modes in six dimensions change. Finally, $p$ specifies on how
many centers the instanton background is located. 
\label{tb:Multiplicities}}
\end{table}

Once the gauge bundle has been topologically characterized, its
embedding as a subgroup $H \in$ SO(32) gauge group has been specified,
and the resulting unbroken gauge group $G$ has been determined, we can
compute the full spectrum using index theorems or equivalently from
the anomaly polynomial of the ten dimensional gaugino. For this we
need to specify the branching of the adjoint representation into a
sum of tensor product representations as 
\equ{
\rep{Ad} ~=~ 496 ~=~ \bigoplus\limits_{\gk} (\rep{r}_\gk, \rep{R}_\gk)~, 
}
where $\rep{r}_\gk$ and $\rep{R}_\gk$ denote irreducible
representations of the non--Abelian part of $H$ and $G$, respectively.
Under the assumption $V_1 = - V_2 = V$, we find that
the multiplicity $N_\gk$ of a state $\rep{R}_\gk$ is determined by
\equ{
N_\gk ~=~ \frac 12\, \frac 1{(2\gp)^2} \int \Big\{
\frac 12 \, \tr_{\rep{r}_\gk} (\cF')^2 
\,+\, \frac 12\, \dim \rep{r}_\gk \, \Big( 
\cF_V^2|_{\rep{R}_\gk} - \frac 1{12}\, \tr \cR^2
\Big) 
\Big\}~, 
\labl{multipop}
}
where $\cF'$ denotes the non--Abelian instanton background, $\cF_V$
the Abelian gauge flux, and $\cR$ the SU(2) curvature two--form. The
integrals over the curvature and the U(1) background follow directly
from the results of earlier parts of this paper, i.e.\ 
\equ{
\frac 1{8\gp^2} \int \tr \cR^2 ~=~ \frac 83~, 
\qquad 
\frac 1{8\gp^2} \int \tr \cF_V^2 ~=~ \frac 13 \, H_V^2~, 
}
where we denote by $H_V = V_I \, H_I$ the Cartan generator of the
Abelian--bundle. The value of the operator $H_V^2$ has to be evaluated
on each of the irreducible representations $\rep{R}_\gk$ as the
multiplicity number~\eqref{multipop} indicates. More care needs to be
taken when computing the integral over the non--Abelian instanton
background $\cF'$, as it also depends on over which 
representation $\rep{r}_\gk$ the trace is taken. For the spin--$\frac 12$
instantons this can be the singlet $\rep{1}$, the fundamental
$\rep{2}$, or the adjoint $\rep{3}$, representations of SU(2); for the
spin-1 instantons only the singlet or triplet representations of SO(3)
are relevant for our purposes. In the cases where there are multiple 
non--Abelian
instantons embedded, also traces over product representations occur. 
The basic values of the possible instanton numbers have been collected
in table~\ref{tb:Multiplicities}.

The resulting spectra are given in the last column of
Table~\ref{classnonab1}. The computation of these spectra requires
mostly standard group theory, see e.g.~\cite{Slansky:1981yr}. As only
the representation theory of the Sp($2n$) groups might be less known,
we have  collected some relevant facts in
Appendix~\ref{sc:spgroups}. The spectra for the pure line bundle
models agree with those given in Ref.~\cite{Honecker:2006qz}; 
the other spectra are novel except that of the standard
embedding.

\begin{table}
%\hspace{-15pt}
\footnotesize
\begin{center} 
\begin{tabular}{|c|c|l|l|}
\hline&&&\\[-2ex]
%%%% Line 1
$( n_1^{1}, n_3^{\frac12}, n_2^{\frac12}, n_1^{\frac12})$ & 
$V_1 = -V_2$ & 
Non--Abelian gauge group & Matter spectrum (up to singlets) 
\\[1ex]\hline\hline &&&\\[-2ex]
%%%% Line 2
%$U_1^3$ 
% $3\alpha$ &
$(2, 0, 0, 0)$ & 
$(0^{16})$ & 
SO(26)
& 
$2 \, (\rep{26})$ 
\\[1ex]\hline&&&\\[-2ex]
$(1,0,0,2)$ & 
$(0^{16})$ &
SO(21) $\times$ Sp(4) 
&
$ \frac 19 \, (\rep{21},\rep{4}) \,+\, 
(\rep{21},\rep{1}) \,+\, (\rep{1},\rep{5}) \,+\,
 3\, (\rep{1},\rep{4})
$
\\[1ex]\hline&&&\\[-2ex]
%%%% Line 3
%/$U_3^2$
% SE &
$(0,1, 0, 0)$ & 
$(0^{16})$ & 
SO(28) $\times$ SU(2)
&
$\frac {10}9\, (\rep{28},\rep{2})$ 
\\[1ex]\hline&&&\\[-2ex]
%%%% Line 4
%$(U_1^2)^4$ 
%$3{\beta}$ & 
$(0, 0,0, 4)$  & 
$(0^{16})$ & 
SO(16) $\times$ Sp(8)
&
$\frac 19\, (\rep{16},\rep{8}) \,+\, (\rep{1},\rep{27})$
%%%% Line 5
\\[1ex]\hline\hline &&&\\[-2ex]
%$(U_1^2)^3$ 
%$3{\ge}
%\left\{\begin{tabular}{c}
%%3{\ge}1\\
%%3{\ge}2\\
%3{\ge}1\\3{\ge}2^*
%\end{tabular}
%\right.$ &
$(1,0,0,1)$ &
$(\overbrace{1,\sm 1}\,,0^{14})$ 
& 
SO(25) 
&
$\frac {17}9\, (\rep{25})$ 
\\[0.4ex]
&
$(1^2,0^{14})$ 
&
SO(21) $\times$ SU(2)$^2$ 
&
$\frac 19\, (\rep{21},\rep{2},\rep{1}) \,+\, 
\frac 19\, (\rep{21},\rep{1},\rep{2}) \,+\, 
 (\rep{21},\rep{1},\rep{1}) \,+\, 
\frac 89\, (\rep{1},\rep{2},\rep{2}) $
\\[0.4ex]
&&&
$\,+\, 3\, (\rep{1},\rep{2},\rep{1}) \,+\, 
3\, (\rep{1},\rep{1},\rep{2})$ 
\\[1ex]\hline&&&\\[-2ex]
$(1, 0,0,0)$ & 
$(2^2,0^{14})$ 
& 
SO(27) 
&
$\frac {19}9\, (\rep{27})$ 
\\[0.4ex]
&
$(1^4,0^{12})$ 
&
SO(21) $\times$ SU(4) 
& 
$\frac 19\, (\rep{21},\rep{4}) 
\,+\, \frac {10}9\, (\rep{1},\rep{6})
\,+\, 3\, (\rep{1},\rep{4}) 
\,+\, (\rep{21},\rep{1}) $
\\[1ex]\hline&&&\\[-2ex]
$(0, 0,0,3)$ & 
$(\overbrace{1,\sm 1}\,,0^{14})$ 
& 
SO(20) $\times$ Sp(4) 
&
$ \frac 19 \, (\rep{20},\rep{4}) \,+\, 
 \frac 89 \, (\rep{20},\rep{1}) \,+\, 
  \frac {28}9 \, (\rep{1},\rep{4}) \,+\, 
 (\rep{1},\rep{5}) $
\\[0.4ex]
&
$(1^2,0^{14})$ 
&
SO(16) $\times$ Sp(6) $\times$ SU(2) 
& 
$ \frac 19 \, (\rep{16},\rep{1}, \rep{2}) \,+\, 
\frac 19\,  (\rep{16},\rep{6}, \rep{1}) \,+\, 
\frac 89 \, (\rep{1},\rep{6}, \rep{2}) \,+\, 
(\rep{1},\rep{14}, \rep{1})
$
%%%% Line 8
\\[1ex]\hline&&&\\[-2ex]
%$(U_1^2)^2$ 
%$3{\eta}
%\left\{\begin{array}{c}
%%3{\eta}1^*\\
%3{\eta}1^*\\3{\eta}2\\3{\eta}3\\
%%3{\eta}5\\
%3{\eta}4^*\\3{\eta}5^*
%\end{array}
%\right.$&
$(0,0,0,2)$ & 
$(2,0^{15})$ 
&
SO(22) $\times$ Sp(4) 
& 
$\frac 19\, (\rep{22},\rep{4}) 
\,+\, \frac {10}9\, (\rep{22},\rep{1}) 
\,+\, (\rep{1},\rep{5}) 
\,+\, \frac {26}9\, (\rep{1},\rep{4}) 
$
\\[1ex] 
& 
$(\overbrace{1^2,\sm 1^2}\,,0^{12})$
&
SO(24) 
& 
$\frac {16}9\, (\rep{24})$ 
\\[0.4ex] 
&
$(\overbrace{1,\sm 1}\,, 1^2,0^{12})$
& 
SO(20) $\times$ SU(2)$^2$
& 
$ \frac 19\, (\rep{20}, \rep{2}, \rep{1}) \,+\, 
 \frac 19\, (\rep{20}, \rep{1}, \rep{2}) \,+\, 
 \frac 89\, (\rep{20}, \rep{1}, \rep{1}) \,+\, 
 \frac 89\, (\rep{1}, \rep{2}, \rep{2}) $
\\[0.4ex] 
&&&
$\,+\,  \frac {28}9\, (\rep{1}, \rep{1}, \rep{2}) \,+\, 
\frac {28}9\, (\rep{1}, \rep{2}, \rep{1})$
\\[0.4ex] 
&
$(1^4,0^{12})$ 
& 
SO(16) $\times$ Sp(4) $\times$ SU(4) 
& 
$ \frac 19\, (\rep{16},\rep{1},\rep{4}) \,+\, 
 \frac 19\, (\rep{16},\rep{4},\rep{1}) \,+\, 
 \frac {10}9\, (\rep{1},\rep{1},\rep{6}) \,+\, 
 \frac 89\, (\rep{1},\rep{4},\rep{4}) $
\\[0.4ex] &&&
$\,+\,
 (\rep{1},\rep{5},\rep{1}) $
\\[1ex]
&
$\frac{1}{2}(\overbrace{1^2,\sm 1^2},1^{12})$
&
SU(12) 
& 
$ \frac {20}9\, (\rep{12}) \,+\, \frac 19\, (\rep{66})$ 
%%%% Line 9
\\[1ex]\hline&&&\\[-2ex]
%$U_1^2$ 
%$3{\gth}
%\left\{\begin{array}{c}
%3{\gth}1^*\\3{\gth}2^*\\3{\gth}3\\3{\gth}4^*\\3{\gth}5^*
%\end{array}
%\right.$ &
$(0,0,0,1)$ 
& 
$(2,\overbrace{1,\sm 1}\,,0^{13})$
&
SO(26) 
& 
$2\, (\rep{26})$ 
\\[0.4ex] 
&
$(2,1^2,0^{13})$
& 
SO(22) $\times$ SU(2)$^2$ 
& 
$\frac 19\, (\rep{22},\rep{2},\rep{1}) 
\,+\, \frac 19\, (\rep{22},\rep{1},\rep{2})
\,+\, \frac {10}9\, (\rep{22},\rep{1},\rep{1})
\,+\, \frac 89\, (\rep{1},\rep{2},\rep{2})  $ 
\\[0.4ex]
&&&
$\,+\, \frac {26}9\, (\rep{1},\rep{2},\rep{1})
\,+\, \frac {26}9\, (\rep{1},\rep{1},\rep{2})$
\\[1ex] 
&
$(\overbrace{1,\sm 1}\,,1^4,0^{10})$
&
SO(20) $\times$  U(4) 
& 
$\frac 19\, (\rep{20},\rep{4}) \,+\, \frac 89\, (\rep{20},\rep{1}) 
\,+\, \frac {28}9\, (\rep{1},\rep{4}) 
\,+\, \frac {10}9\, (\rep{1},\rep{6})$ 
\\[0.4ex] 
&
$(1^6,0^{10})$ 
&
SO(16)  $\times$ SU(6) $\times$ SU(2)
& 
$\frac 19\, (\rep{16},\rep{6},\rep{1}) 
\,+\, \frac19\,(\rep{16},\rep{1},\rep{2})
\,+\, \frac 89\, (\rep{1},\rep{6},\rep{2}) 
\,+\, \frac {10}9\, (\rep{1},\rep{15},\rep{1})$ 
\\[1ex]
&
$\frac{1}{2}(\overbrace{1,\sm 1},1^{13},3)$ 
& 
SU(13) 
& 
$\frac {21}9\, (\rep{13}) \,+\, \frac 19\, (\rep{78})$ 
%%%% Line 10
\\[1ex]\hline\hline&&&\\[-2ex]
%$3\lambda$ & 
$(0,0,0,0)$ &  
$(2^2,0^{14}) $
& 
SO(28) $\times$ SU(2) 
& 
$\frac{10}9\, (\rep{28},\rep{2})
%\,+\, \frac{46}9\, (\rep{1})
$ 
\\[1ex] 
&
$(2,1^4,0^{11})$
& 
SO(22) $\times$ SU(4) 
& 
$\frac 19\, (\rep{22},\rep{4}) \,+\, \frac {10}9\, (\rep{22},\rep{1}) 
\,+\, \frac{10}9\, (\rep{1},\rep{6}) 
\,+\, \frac{26}{9}\, (\rep{1},\rep{4})$
\\[1ex] 
&
$(1^8,0^{8})$
&
SO(16) $\times$ SU(8) 
& 
$\frac 19\, (\rep{16},\rep{8}) \,+\, \frac{10}9\, (\rep{1},\rep{28})$
\\[1ex] 
&
$\frac{1}{2}(1^{14},3^{2})$ 
&
SU(14) $\times$ SU(2) 
& 
$\frac 19\, (\rep{91},\rep{1}) \,+\, \frac{11}9\,(\rep{14},\rep{2}) 
%\,+\, \frac{25}9\,(\rep{1})
$
\\[1ex]\hline
\end{tabular}
\end{center} 
\caption{\label{classnonab1} 
This table gives the Abelian and non--Abelian bundles fulfilling the
local Bianchi identity~\eqref{localBI} on the resolved
$\Cplx^2/\Intr_3$ singularity, 
and the resulting models. The first column indicates the instanton
numbers of the non--Abelian bundles. The second column gives the
Abelian bundle vectors. The overbrace indicates the amount of
``overlap'', i.e.\ shared Cartan generators, there is between Abelian 
background and the SU(2)
instanton(s). The third column lists the possible unbroken
non--Abelian gauge group. The final column gives the resulting 
spectrum up to singlets. 
}
\end{table}

\subsection{Supersymmetric blowups}

\begin{table}
%\hspace{-15pt}
\footnotesize
\begin{center} 
\begin{tabular}{|c|c|l|lll|}
\hline&&&&&\\[-2ex]
%%%% Line 1
$( n_1^{1}, n_3^{\frac12}, n_2^{\frac12}, n_1^{\frac12})$ & 
$V_1 = -V_2$ & 
Unbroken gauge group & \# & Twisted  & Untwisted
\\[1ex]\hline\hline &&&&&\\[-2ex]
%%%% Line 2
%$U_1^3$ 
% $3\alpha$ &
$(2, 0, 0, 0)$ & 
$(0^{16})$ & 
SO(26) %$\times$ U(1)  
& 3a & $(\rep{28},\rep{2}), (\rep{1})$ & 
\\[1ex]\hline&&&&&\\[-2ex]
$(1,0,0,2)$ & 
$(0^{16})$ &
SO(21) $\times$ Sp(4) 
& 3b & $(\rep{22},\rep{1}), (\rep{1},\rep{10})$ 
& $(\rep{22},\rep{5})$
\\[1ex]\hline&&&&&\\[-2ex]
%%%% Line 3
%/$U_3^2$
% SE &
$(0,1, 0, 0)$ & 
$(0^{16})$ & 
SO(28) $\times$ SU(2)
&
3a 
&
$2\times (\rep{1})$
&
\\[1ex]\hline&&&&&\\[-2ex]
%%%% Line 4
%$(U_1^2)^4$ 
%$3{\beta}$ & 
$(0, 0,0, 4)$  & 
$(0^{16})$ & 
SO(16) $\times$ Sp(8)
&
3c &  $(\rep{1},\rep{28}), (\rep{1})$ &
\\[0.4ex]
&&&& $(\rep{1},\rep{28}), (\rep{1})$ & $(\rep{1},\rep{28})$ 
%%%% Line 5
\\[1ex]\hline\hline &&&&&\\[-2ex]
%$(U_1^2)^3$ 
%$3{\ge}
%\left\{\begin{tabular}{c}
%%3{\ge}1\\
%%3{\ge}2\\
%3{\ge}1\\3{\ge}2^*
%\end{tabular}
%\right.$ &
$(1, 0,0,1)$ & 
$(\overbrace{1,\sm 1}\,,0^{14})$ 
& 
SO(25)
&3a & $(\rep{28},\rep{2}), (\rep{1})$ 
& $(\rep{28},\rep{2})$ 
\\[0.4ex]
&
$(1^2,0^{14})$ 
&
SO(21) $\times$ SU(2) $\times$ SU(2) 
& 3b & $(\rep{22},\rep{1}), (\rep{1},\rep{10}), (\rep{1},\rep{5})$  
& $(\rep{1},\rep{10})$ 
%%%% Line 8
\\[1ex]\hline&&&&&\\[-2ex]
$(1, 0,0,0)$ & 
$(2,0^{15})$ 
& 
SO(27) %$\times$ U(1) 
& 3a & $(\rep{28},\rep{2}), (\rep{1})$ 
&  $(\rep{28},\rep{2})$ 
\\[0.4ex]
&
$(1^4,0^{12})$ 
&
SO(21) $\times$ SU(4) %$\times$ U(1) 
& 3b  & $(\rep{22},\rep{1}), (\rep{1},\rep{5})$  
& $(\rep{22},\rep{5})$ 
%%%% Line 8
\\[1ex]\hline&&&&&\\[-2ex]

$(0, 0,0,3)$ & 
$(\overbrace{1,\sm 1}\,,0^{14})$ 
& 
SO(20) $\times$ Sp(4) %$\times$ U(1) 
& 3b & $(\rep{22},\rep{1}), (\rep{1},\rep{10})$ 
& $(\rep{22},\rep{5})$ 
\\[0.4ex]
&
$(1^2,0^{14})$ 
&
SO(16) $\times$ Sp(6) $\times$ SU(2) %$\times$ U(1)
& 3c & $(\rep{1},\rep{28}), (\rep{1})$  & $(\rep{1},\rep{28})$ 
%%%% Line 8
\\[1ex]\hline&&&&&\\[-2ex]
%$(U_1^2)^2$ 
%$3{\eta}
%\left\{\begin{array}{c}
%%3{\eta}1^*\\
%3{\eta}1^*\\3{\eta}2\\3{\eta}3\\
%%3{\eta}5\\
%3{\eta}4^*\\3{\eta}5^*
%\end{array}
%\right.$&
$(0,0,0,2)$ & 
$(2,0^{15})$ 
&
SO(22) $\times$ Sp(4) %$\times$ U(1) 
& 3b & $(\rep{1},\rep{10}), (\rep{1},\rep{5})$ &
\\[0.4ex]
&&&&    $(\rep{1},\rep{10}), (\rep{1},\rep{5})$ &  $(\rep{1},\rep{10})$
\\[1ex] 
& 
$(\overbrace{1^2,\sm 1^2}\,,0^{12})$
&
SO(24) % $\times$  U(1) $\times$ U(1) 
& 3a & $(\rep{28},\rep{2}), (\rep{1})$ & $(\rep{28},\rep{2})$ 
\\[0.4ex] 
&
$(\overbrace{1,\sm 1}\,, 1^2,0^{12})$
& 
SO(20) $\times$ SU(2)$^2$% $\times$ U(1) $\times$ U(1) 
& 3b & $(\rep{22},\rep{1}), (\rep{1},\rep{10})$ 
& $(\rep{22},\rep{5}), (\rep{1}, \rep{10})$ 
\\[0.4ex] 
&
$(1^4,0^{12})$ 
& 
SO(16) $\times$ Sp(4) $\times$ SU(4) %$\times$ U(1)
& 3c & $(\rep{1},\rep{28}),(\rep{1})$  & $(\rep{1},\rep{28})$ 
\\[1ex]
&
$\frac{1}{2}(\overbrace{1^2,\sm 1^2},1^{12})$
&
SU(12)% $\times$ U(1) $\times$ U(1)
& 
3e & $(\rep{14},\rep{2},\rep{1}), (\rep{2},\rep{1},\rep{1}), (\rep{1})$ 
&  $(\rep{14},\rep{2},\rep{2})$
%%%% Line 9
\\[1ex]\hline&&&&&\\[-2ex]
%$U_1^2$ 
%$3{\gth}
%\left\{\begin{array}{c}
%3{\gth}1^*\\3{\gth}2^*\\3{\gth}3\\3{\gth}4^*\\3{\gth}5^*
%\end{array}
%\right.$ &
$(0,0,0,1)$ 
& 
$(2,\overbrace{1,\sm 1}\,,0^{13})$
&
SO(26) %$\times$ U(1) $\times$ U(1) 
& 3a & $(\rep{28},\rep{2}), (\rep{1})$ & 
\\[0.4ex] 
&
$(2,1^2,0^{13})$
& 
SO(22) $\times$ SU(2)$^2$ %$\times$ U(1) $\times$ U(1) 
& 3b & $(\rep{1},\rep{10}), (\rep{1},\rep{5})$ &  $(\rep{1},\rep{10})$
\\[1ex] 
&
$(\overbrace{1,\sm 1}\,,1^4,0^{10})$
&
SO(20) $\times$  U(4) %$\times$ U(1) 
& 3b &  $(\rep{22},\rep{1}), (\rep{1},\rep{5})$ & $(\rep{22},\rep{5})$
\\[0.4ex] 
&
$(1^6,0^{10})$ 
&
SO(16) $\times$ SU(2) $\times$ SU(6)% $\times$ U(1) 
& 3c & $(\rep{1},\rep{28}), (\rep{1})$ & $(\rep{1},\rep{28})$ 
\\[1ex]
&
$\frac{1}{2}(\overbrace{1,\sm 1},1^{13},3)$ 
& 
SU(13) %$\times$ U(1) $\times$ U(1)
& 3e &  $(\rep{14},\rep{2},\rep{1}), (\rep{2},\rep{1},\rep{1}), (\rep{1})$ 
&  $(\rep{14},\rep{2},\rep{2})$
%%%% Line 10
\\[1ex]\hline\hline&&&&&\\[-2ex]
%$3\lambda$ & 
$(0,0,0,0)$ &  
$(2^2,0^{14}) $
& 
SO(28) $\times$ SU(2)% $\times$ U(1)  
& 3a & $2 \times (\rep{1})$ & 
\\[1ex] 
&
$(2,1^4,0^{11})$
& 
SO(22) $\times$ SU(4)% $\times$ U(1) $\times$ U(1) 
& 3b & $2 \times (\rep{1},\rep{5})$ & 
\\[1ex] 
&
$(1^8,0^{8})$
&
SO(16) $\times$ SU(8)% $\times$ U(1)
& 3c & $2 \times (\rep{1})$ & 
\\[1ex] 
&
$\frac{1}{2}(1^{14},3^{2})$ 
&
SU(14) $\times$ SU(2) %$\times$ U(1) $\times$ U(1)
& 3e & $2 \times (\rep{1},\rep{1},\rep{2})$ & 
\\[1ex]\hline
\end{tabular}
\end{center} 
\caption{\label{classnonab2} 
The first three columns contain the same information as
Table~\ref{classnonab1}. The final columns indicate from which of five
heterotic $\Intr_3$ models, listed in Table~\ref{tb:HetC2Z3}, these bundle
models can be obtained by switching on VEV's for the indicated
twisted and untwisted states.}
\end{table}

We study blow-ups of the $\Intr_3$ heterotic orbifold models, that
preserve six dimensional supersymmetry by switching on VEV's for 
twisted and possibly also untwisted states, and that can be identified
with the smooth bundle models listed in Table~\ref{classnonab1}.  
The analysis can be performed entirely at the classical level, because
in six dimensional super--Yang--Mills theory dangerous loop corrections 
to the potential are absent. (This is of course unlike the four dimensional 
case, where one-loop  Fayet--Iliopoulos corrections may arise.)

The study of flat directions of the potential $V$ involves the three
real auxiliary fields, $D^i_a$ with $i=1,2,3$, of six dimensional
super Yang--Mills theory  
\equ{
V ~=~ \frac 12\, \sum_{i,a} (D^i_a)^2~, 
\qquad 
D^i_a ~=~ \gs^i_{\ga\gb}\,  \phi_\ga^\dag \, T_{a}\phi_\gb~, 
}
where the representation indices on the complex scalar components
$\phi_1, \phi_2$ of a hypermultiplet and gauge generator $T_a$
have been suppressed. It turns out convenient to use four dimensional 
$\cN=1$ notation of a real D--term $D_a=D_a^3$ and a complex F--term 
$F_a= (D_a^1+iD_a^2)/\sqrt 2$. Since the complex scalar $\gf_1$ and
$\gf_2$ components of hypermultiplets are in complex
conjugate representations, we have 
\equ{
V ~=~ \frac 12 \sum_a D_a^2 \,+\, \sum_a \bF_a F_a~, 
\qquad 
D_a ~=~ \bgf_1 T_a \gf_1 \,-\, \gf_2 T_a \bgf_2~, 
\qquad 
\bF_a ~=~ \gf_2 T_a \gf_1~. 
}
Therefore, if the scalars in the hypermultiplet are internally
aligned, i.e.\ 
\equ{
\gf_2 ~=~ \ga_\gf\, \bgf_1~, 
\quad |\ga_\gf| ~=~ 1
\quad \Ra \quad 
\bF ~=~ \ga_\gf\, \bgf_1 T_a \gf_1~, 
}
the D--term vanishes immediately, and the F--term takes the form 
of a D--term but with a phase $\ga_\gf$ as pre--factor. If one has more
than one hypermultiplet, the alignment can happen in each hyper
multiplet separately, which gives a collection of phases, and relative
signs in particular. This simplifies the analysis considerably: One
does not have to worry anymore about D--terms and the phases may be
used to make the F--terms vanish as well.

The subsequent analysis of the flat directions is straightforward but
somewhat tedious. 
We have diverted most of this discussion 
to Appendix~\ref{sc:flatness}; here we only summarize the
results of the complete analysis in Table~\ref{classnonab2}. In this
table we list for each of the bundle models given in
Table~\ref{classnonab1} from which heterotic orbifold models,
classified in Table~\ref{tb:HetC2Z3}, it can be obtained by switching on
the VEV's for the hypermultiplets listed in the last column of
Table~\ref{classnonab2}. This table shows that each bundle model, for
which explicit solutions to the Hermitean Yang--Mills equations
exist, indeed corresponds to an F-- and D--flat direction.

One can follow this correspondence of the bundle models and the
orbifold models also at the level of the spectra. We have checked,
that the non--Abelian spectra of the orbifold models, with appropriate
VEV's switched on, results in branching of the matter representation
giving precisely the non--Abelian 
spectra of the bundle models. This identification is exact if one
takes Higgsing of vector multiplets due to symmetry breaking into
account, that eats away some hypermultiplets. Because of six
dimensional chirality, states can only pair up and become massive,
provided that one is a vector multiplet and the other a hypermultiplet. This means that the index theorem exactly determines the
number of massless hyper (including non--Abelian singlets) and vector
multiplets. In Table~\ref{classnonab1} we refrained from giving the
multiplicities of singlet states; they can either be directly computed
via the index theorem, or using the fact that the pure gravitational
anomaly gives a relation between the number of vector multiplets and
hypermultiplets~\cite{Green:1984bx,Erler:1993zy}.

From Table~\ref{classnonab2} we can determine some relations between
the VEV's of twisted and untwisted states of the orbifold model and
the corresponding bundle model. In particular, we see that all pure
line bundle models are obtained by switching on VEV's for two identical twisted
hypers. All other bundle models have different hypermultiplets
switched on; except for the standard embedding model with $n=(0,1,0,0)$.

\subsection{Modified local Bianchi identity}
\labl{sc:bianchi}

The comparison between possible VEV configurations of
orbifold models and the explicit bundle models constructed here,
indicates that our list of bundles is not complete: There are
supersymmetric VEV
assignments that do not seem to have a counter part as a bundle
model. Before we explain what is going on here, we first give two
examples of this situation:

First of all, notice that all the constructed bundle
models are obtained by switching on VEV's in orbifold models 3a, 3b,
3c and 3e of Table~\ref{tb:HetC2Z3}, while model 3d is never
used. Nevertheless this model has a fully flat direction with
simultaneously suitably aligned VEV's of the $(\rep{10},\rep{11})_1$,
$(\rep{1},\rep{55})_2$ and $(\rep{1},\rep{11})_{-8/3}$ breaking the
gauge group to SO(9)$\times$Sp(10).

A second example is provided by orbifold model 3c. We see from
Table~\ref{classnonab2} that all the bundle models with an SO(16)
group factor result from this orbifold model by switching on VEV's of
one or two twisted singlets $(\rep{1})_{8/3}$ and the twisted and
untwisted anti--symmetric tensors, $(\rep{1},\rep{28})_{\sm 2/3}$ and
$(\rep{1},\rep{28})_{1}$. Following the analysis of
Appendix~\ref{sc:flatness} one concludes that with these multiplets
taking VEV's the possible unbroken gauge group could be any of the
ones listed in Table~\ref{tb:missingmodels}. 
These different possibilities arise because of the VEV's for the
anti--symmetric tensors: They can be skew--diagonalized. Then
depending on whether some or all of its diagonal entries are equal and
/ or zero, one of the above mentioned gauge groups is realized. (For
example: All entries zero gives SU(8), all entries equal but non--zero
gives Sp(8), and finally all entries different gives SU(2)$^4$.)
Table~\ref{classnonab1} does not contain the gauge group factors 
Sp(4)$\times$Sp(4), SU(4)$\times$SU(2)$^2$, Sp(4)$\times$SU(2)$^2$ and
SU(2)$^4$, hence there are bundle models missing.

As a side remark we note, that this example also shows that many
different bundle models, characterized by different topological
parameters are actually related to each other by continuous
deformations of the VEV's of twisted and untwisted states of the
corresponding orbifold model. The precise relation between the moduli
space of VEV configurations and bundle models is beyond the scope of
this paper. Presumably this requires to analyze the full gauge bundle
moduli space using the ADHM~\cite{Atiyah:1978ri,Dorey:2002ik} and
Kronheimer--Nakajima~\cite{Kronheimer:1989zs,Kronheimer:1990,Nakajima:1990}
constructions, see e.g.~\cite{Douglas:1996sw,Bianchi:1996zj,Douglas:1997de,Fucito:2004ry}.

Bundle realizations of these and other VEV configurations of orbifold
models can be obtained realizing that the {\em local} Bianchi
identity~\eqref{localBI} is a sufficient condition to uncover consistent
models but certainly not a necessary condition. Indeed, only on a
compact K3 the integrated Bianchi identity needs to vanish. This means
that if one has a compact orbifold, say like $T^4/\Intr_3$, the sum of
the instanton numbers from all fixed points needs to equal 24. The
local Bianchi identity~\eqref{localBI} is obtained by splitting up the
total instanton number of K3 equally over all 9 fixed points of
$T^4/\Intr_3$. The total instanton number 24 cannot be completely
arbitrarily distributed over the various fixed points, since the
instanton number is quantized
itself~\cite{Seiberg:1996vs,Duff:1996rs,Aldazabal:1997wi}: The basic
unit of instanton number that can be moved around equals 1.  
This means that the local Bianchi identity~\eqref{localBI} equals $K =
8/3$ mod $1$.  
The additional constraint, that the first Chern class of the bundle is
even, implies that $K = 8/3$ mod $2$, unless $n_2^{1/2}$ is odd. 
This coincides precisely with the (weak) modular invariance condition
for a local orbifold shift vector.

\begin{table} 
\begin{center}
\tabu{|c|c|c|}{
\hline&&\\[-2ex]
 gauge group & bundle realization & $K$ 
\\[1ex]\hline\hline &&\\[-2ex]
SO(16)$\times$SU(8) & $V_1=(1^8,0^8)$ & $\frac83$ 
\\[1ex]%\hline &&&\\[-2ex]
SO(16)$\times$Sp(8)& $n_1^{1/2} = 4$ & $\frac83$ 
\\[1ex]%\hline &&&\\[-2ex]
SO(16)$\times$Sp(6)$\times$SU(2)& $n_1^{1/2}=3,V_1=(1^2,0^{14})$ & $\frac83$ 
\\[1ex]%\hline &&&\\[-2ex]
SO(16)$\times$SU(4)$\times$Sp(4) & $n_1^{1/2}=2,V_1=(1^4,0^{12})$ & $\frac83$ 
\\[1ex]\hline\hline &&\\[-2ex]
SO(16)$\times$SU(6)$\times$SU(2)& $n_1^{1/2}=1, V_1=(1^6,0^8)$ & $\frac83$ 
\\[1ex]%\hline &&&\\[-2ex]
SO(16)$\times$Sp(4)$\times$Sp(4) & $n_2^{1/2} = n_1^{1/2} = 2$ 
&  $\frac{14}3$
\\[1ex]%\hline &&&\\[-2ex]
SO(16)$\times$SU(4)$\times$SU(2)$^2$ & 
$n_1^{1/2}=1, V_1 = (2^2, 1^4, 0^{10})$ &  $\frac{14}3$
\\[1ex]%\hline &&&\\[-2ex]
SO(16)$\times$Sp(4)$\times$SU(2)$^2$ & 
$n_1^{1/2} =2, V_1=(2^2, 1^2,0^{12})$ &  $\frac{14}3$
%\\[1ex]%\hline &&&\\[-2ex]
\\[1ex]\hline\hline &&\\[-2ex]
SO(16)$\times$SU(2)$^4$ & 
$n_3^{1/2} = n_1^{1/2} =1, V_1 = (2^2, 1^2,0^{12})$ &  $\frac{20}3$
 \\[1ex]\hline
}
\end{center}
\caption{\label{tb:missingmodels}
We give possible bundle realizations of all VEV configurations of 
model 3c. The last column indicates for which models a modification of
the local Bianchi identity is required, i.e.\ $K \neq 8/3$. (Only the
non--vanishing gauge instanton numbers are given, and $V_1= - V_2$ is
assumed.) 
}
\end{table}

The smallest local Bianchi identity has $K= 2/3$ in~\eqref{localBI}. 
There are two solutions to this equation: i) $n_1^{1/2}=1$,
$V_1=V_2=0$, which results in the unbroken gauge group
$\SO{28}\times\SU{2}$, and ii) $n_1^{1/2}=0$, $V_1=-V_2=(1^2,0^{14})$,
with unbroken gauge group $\SO{28}\times \SU{2}$. Thus both are 
VEV configurations of orbifold model 3a, which we had already found.

Using the modified Bianchi identity, eq.~\eqref{localBI} for an
instanton number $K = 14/3$ a bundle realization of the VEV
configuration of model  3d can be found: The bundle is characterized
by $n_1^1=1$ and $n_1^{1/2} =5$. Also the VEV configurations of 3c
with gauge groups Sp(4)$\times$Sp(4), SU(4)$\times$SU(2)$^2$,
Sp(4)$\times$SU(2)$^2$ and SU(2)$^4$, discussed above, can be
identified. For each of these models we give a bundle candidate in
Table~\ref{tb:missingmodels}. To compute the spectra of 
these models is challenging because for that we need a modified index
theorem that takes the non--vanishing three--form flux $H_3$ into
account. Indeed, using the standard index theorem ensures an
anomaly--free spectrum only in case the Bianchi identity is fulfilled
\cite{Witten:1984dg,Green:1984bx}.

\section{Conclusions and outlook}
\labl{sc:concl}

The construction of stable non--Abelian bundles on Calabi--Yau
manifolds is one of the outstanding problems in both mathematics and
theoretical physics. Yet to determine the full phenomenological
potential of heterotic string constructions this is of fundamental
importance. In this paper we exploited the fact that
well--known instantons on Eguchi--Hanson spaces provide explicit
examples of stable bundles on non--compact four dimensional
$\Cplx^2/\Intr_n$ orbifold blowups with non--Abelian structure groups. 
Because in addition also line bundles have been constructed on
Eguchi--Hanson spaces explicitly, we have access to a substantial
class of bundles that can be used for six dimensional model building. 
Using this we gave a complete classification of all possible
combinations of these instantons with Abelian gauge fluxes, that
fulfill the local Bianchi identity constraint on the $\Cplx^2/\Intr_3$ 
resolution. Spectra were computed using index theorems; to obtain
anomaly--free spectra it was crucial that the Bianchi identity was
fulfilled locally. The resulting effective six dimensional models have been
listed in Table~\ref{classnonab1}.

All of these gauge backgrounds can be related to a configurations of
VEV's of states present in 
the corresponding heterotic orbifold models. 
For models with only Abelian gauge fluxes always two
identical twisted hypermultiplets take VEV's, confirming our
previous findings~\cite{Honecker:2006qz,Nibbelink:2007rd,Nibbelink:2007pn}, 
that line bundle models correspond to orbifold models with a single
twisted VEV switched on. 
For non--Abelian gauge fluxes or gauge backgrounds that combine both
line bundles and bundles with non--Abelian structure groups, we always
need combinations of simultaneous VEV's of twisted and often even
untwisted states to identify matching orbifold constructions.  
In all cases we confirmed that both the gauge groups and
spectra are identical in the orbifold and bundle perspectives. The
multiplicities of states in the smooth construction, computed using the
index theorem on local resolutions, seems to take rather arbitrary
values, given in the final column of Table~\ref{classnonab1}. All
these values can be understood from the orbifold perspective as the
combination of twisted states with integral multiplicities, untwisted
states with multiplicity $1/9$ (because they are bulk modes shared
between nine orbifold fixed points), and the effect of Higgsings
that take away multiples of $1/9$. Therefore, this provides
stringent consistency checks on our results.

We have shown that each combination of instantons and Abelian gauge
fluxes that fulfill the local Bianchi identity corresponds to a VEV
configuration of a certain heterotic orbifold. One may wonder whether
one can reverse the statement: Each supersymmetric system of
VEV's correspond to a configuration of instantons and gauge
fluxes. Presumably this statement is true, but certainly not all these
configurations satisfy the local Bianchi identities. Indeed, we
observed that model 3d of Table~\ref{tb:HetC2Z3} is not used at all as
an orbifold realization of a bundle model that satisfies this
condition, see Table~\ref{classnonab2}, even though it 
definitely possesses flat directions. If we give up the local Bianchi
identity and allow that it differs by some instanton units, a
configuration can be identified that leads to the same gauge group as
one obtains from the VEV configuration. To confirm the matching on the
level of the spectra is hampered by the fact, that index theorems on
non--compact spaces cannot be employed when the local Bianchi  is not
satisfied. A generalization of the index theorem in the presence of
the corresponding three form $H$--flux is needed.

The situation is similar for the possible VEV configurations of the
other orbifold models. For concreteness we focused on model 3c: Only 
some of its VEV configurations are realized as bundle models
satisfying the local Bianchi identity. Other VEV assignments can only 
be realized, when it is only fulfilled up to a number of instanton
units. The resulting Bianchi identity is then very similar to the
modular invariance condition of heterotic orbifolds. All these
different bundle models correspond to VEV configurations which are all
continuously connected to each other. Different bundle models often
only correspond to very similar VEV configurations, except
that in one case the VEVs are equal, in the other they are different. 
One does not need to take large numbers of VEVs to zero to
interpolate between such configurations, therefore these 
transitions are deformations of the bundle rather than flops. 
In light of this one may wonder
what the topological classification of the bundles exactly means. The
description of bundles on Eguchi--Hanson spaces employed by us is not
the most general:  The Kronheimer--Nakajima
construction~\cite{Kronheimer:1990} describes the  
full moduli space on such ALE gravitational instantons, and might
therefore be a more appropriate setting for this comparison.

Most of the findings reported in this work relied on the crucial fact
that on Eguchi--Hanson spaces, Abelian gauge backgrounds and
non--Abelian instanton configurations are known. 
Explicit resolutions of $\Cplx^3/\Intr_n$ for $n > 3$ 
orbifolds are not known, hence
to have access to bundles with non--Abelian structure groups on
$\Cplx^3/\Intr_n$ resolutions is much more challenging. (Of course one
always has the standard embedding, but precisely since it immediately
fulfills the local Bianchi identity, it only corresponds to one
configuration.) Yet this is of
great importance because there are certain six dimensional orbifolds,
like the $T^6/\Intr_{6\text{--II}}$ for which a large pool of
MSSM--like models have been constructed recently. Resolutions 
of generic $\Cplx^3/\Intr_n$ orbifolds and their line bundles are only
known in toric geometry. 
In the hope to find a framework that allows us to describe stable
bundles with non--Abelian structure groups on toric resolutions of
such orbifold singularities, we reformulated the description of the
Eguchi--Hanson instantons  in terms of a toric geometry--like
language.

\subsection*{Acknowledgments}

We would like to thank Massimo Bianchi, Kang--Sin Choi, Tae-Won Ha,
Arthur Hebecker, Maximilian Kreuzer, Michael Ratz, Emanuel Scheidegger,
Stefan Vandoren and Jenny Wagner for stimulating discussions and
correspondence. F.P.C. is grateful to the Institut fur Theoretische
Physik, Heidelberg, Germany, for support and hospitality during visits
in the early stage of this work. The work of F.P.C is supported by FCT
through  the grant SFRH/BPD/20667/2004.
The work of MT is supported by the European Community through the
contract N 041273 (Marie Curie Intra-European Fellowships).
He is also  partially supported by the ANR grant  ANR-05-BLAN-0079-02,
the RTN contracts MRTN-CT-2004-005104 and MRTN-CT-2004-503369, 
the CNRS PICS \#~2530, 3059 and 3747, and by the European Union
Excellence Grant  MEXT-CT-2003-509661.

\appendix
\def\theequation{\thesection.\arabic{equation}} 
\setcounter{equation}{0}

\section{Some Sp(${\bf 2n}$) representation theory} 
\labl{sc:spgroups} 
\setcounter{equation}{0}

This Appendix is devoted to some elementary properties of
representation of Sp($2n$) groups and how they arise in branching from
SO($4n$) groups. Sp($2n$) groups are less common in physics, for that
reason we review the properties that we need here. (See for a more
extensive discussion Ref.~\cite{Georgi:1982jb}.) 
The group Sp($2n$) is defined as the group of real
matrices that leave a symplectic form (anti--symmetric $2n\times2n$
matrix) $\gO$ invariant
\equ{
S^T \, \gO\, S ~=~ \gO~,
\qquad 
\gO ~=~ \Id_n\otimes \ge ~=~ \pmtrx{~0 & \Id_n \\ -\Id_n & 0}~. 
}
The form of the symplectic matrix $\gO$ given here can be obtained by
a suitable basis choice. Alternatively one can define this group as
the set of unitary matrices $U \in \SU{2n}$ that leave this
symplectic form invariant $U^\dag \gO U = \gO$. This group is then
also often referred to as USp($2n$), both definitions in fact define
the same abstract group.

We list the basic representations of Sp($2n$). Since Sp($2n$) is
defined as a matrix group, its fundamental representation is the $2n$
component vector representation $\rep{2n}$ on which these matrices act
naturally. The adjoint representation is defined as the algebra of the
group. Writing an algebra element $A$ as a block matrix, we find that
its matrix blocks satisfy 
\equ{
A ~=~ \pmtrx{ \ga & \gb \\ \gg & \gd }~, 
\qquad 
\gb^T ~=~ \gb~, \quad 
\gg^T ~=~ \gg~, \quad 
\gd ~=~ - \ga^T~. 
}
Therefore the adjoint consists of $n(2n+1)$ components in
total. This corresponds to symmetric Hermitian $2n\times 2n$ matrices,
that are the generators of Sp($2n$) as a subgroup of the unitary
group. We can also consider the anti--symmetric Hermitian
matrices. This does not give directly an irreducible representation 
because the symplectic form $\gO$ itself is anti--symmetric. Using it
we can define the traceless anti--symmetric representation 
$\rep{[2n]}_2$ with $n(2n-1)-1$ components. These representations for
Sp($2n$) groups up to $n = 5$ are collected in
Table~\ref{tb:spgroups}.

\begin{table}
\begin{center} 
\tabu{|c||c|c|c|c|c|}{
\hline&&&&&\\[-2ex]
$n$ & 1 & 2 & 3 & 4 & 5 
\\[1ex]\hline &&&&&\\[-2ex]
Sp($2n$) & Sp(2) & Sp(4) & Sp(6) & Sp(8) & Sp(10)
\\[1ex]\hline\hline &&&&&\\[-2ex]
$\rep{Fund} = \rep{2n}$ & $\rep{2}$ &  $\rep{4}$ &  $\rep{6}$ &  $\rep{8}$  &  $\rep{10}$ 
\\[1ex]\hline &&&&&\\[-2ex]
$\rep{Ad} = \rep{n(2n+1)}$ &  $\rep{3}$ &  $\rep{10}$ &  $\rep{21}$ &  $\rep{36}$  &
$\rep{55}$  
\\[1ex]\hline &&&&&\\[-2ex]
$\rep{[2n]}_2 = \rep{n(2n \sm 1) \sm1}$ & - &  $\rep{5}$ &  $\rep{14}$ &  $\rep{27}$  &
$\rep{48}$
 \\[1ex]\hline
}
\end{center}
\caption{\label{tb:spgroups} The elementary representations of the
smallest Sp($2n$) groups are listed. The representations for the
smallest two makes sense in view of the isomorphisms 
$\Sp{2} = \SU{2}$ and $\Sp{4} = \SO{5}$.}
\end{table}

To compute the spectra of models when Sp--groups appear in the main
part of the text the branching of SO($4n$) and SU($2n$) to Sp($2n$)
are crucial. The relevant branching rules read 
\equ{
\arry{ccl}{
\SO{4n} & \ra & \Sp{2n} \times \SU{2}~, 
\\[1ex] 
\rep{4n} & \ra & (\rep{2n}, \rep{2})~, 
\\[1ex] 
\rep{2n(4n-1)} & \ra &  
(\rep{n(2n+1)}, \rep{1}) \,+\,  (\rep{1}, \rep{3}) 
\,+\,  (\rep{n(2n\sm 1)\sm 1}, \rep{3})~, 
}
}
and 
\equ{
\arry{ccl}{
\SU{2n} & \ra & \Sp{2n}~, 
\\[1ex]
\rep{2n} & \ra & \rep{2n}~, 
\\[1ex] 
\rep{n(2n\sm 1)} & \ra & (\rep{n(2n\sm1)\sm1}) \quad+\quad (\rep{1})~, 
\\[1ex]
\rep{4n^2\sm1} & \ra & (\rep{n(2n+1)}) \quad+\quad (\rep{n(2n\sm1)\sm1})~. 
}
}

\section{Clifford algebras for $\boldsymbol{\SO{4N}}$}
\label{sc:clifford}

In the main text we rely at certain points heavily on some properties
of Clifford algebras and spinor representations of $\SO{N}$ groups. A
convenient way of introducing their properties is to make use of an
explicit basis. For the purposes of this paper we make the following
choices. The standard Pauli matrices
\equ{
\gs_1 ~=~ \pmtrx{0 & 1\\[1ex] 1 & 0}~, 
\quad
\gs_2 ~=~ \pmtrx{0 & \sm i \\[1ex] i &  0}~, 
\quad
\gs_3 ~=~ \pmtrx{1 & 0\\[1ex] 0& \sm 1}~, 
}
are defined such that $\gs_1 \gs_2 = i \gs_3$.

The four dimensional Euclidean gamma matrices can be chosen as 
\equ{
\gg_i ~=~ \pmtrx{0 & i \gs_i \\[1ex] \sm i \gs_i & 0}~, 
\qquad 
\gg_4 ~=~ \pmtrx{0 & \Id_2 \\[1ex] \Id_2 & 0}~.
}
The spin generators $\frac 12 \gg_{AB} = \frac 14[\gg_A,\gg_B]$ of
$\SO{4}$ are then given by  
\equ{
\sfrac 12\, \gg_{ij} ~=~ \sfrac i2 \, \ge_{ijk}\, 
\pmtrx{\gs_k & 0\\[1ex] 0 & \gs_k}~,
\qquad 
\sfrac 12\, \gg_{k4} ~=~ \sfrac i2 \,
\pmtrx{\gs_k & 0\\[1ex] 0 & \sm \gs_k}~.
}
The product of all four gamma matrices defines the chirality operator
$\gg=\gg_1\gg_2\gg_3\gg_4$. Using it one defines the chiral
projections of the spin generators 
\equ{
\gg_{AB}^\pm ~=~ \gg_{AB} \, P^\pm~, 
\qquad 
P^\pm ~=~ \frac {1 \,\pm\, \gg}2~. 
}
Notice that 
$\gg \gg_{AB} = - \frac 12 \ge_{ABCD}\, \gg_{CD}$, hence positive
chirality corresponds to self--duality, see~\eqref{hodge} in our
conventions.

The eight dimensional Euclidean Clifford algebra and spin group are
obtained straightforwardly from the four dimensional one. We define 
\equ{
\gG_A ~=~ \gg_A\otimes \Id~, 
\qquad 
\gG_{4+A} ~=~ \gg \otimes \gg_A~,
}
as the basis of the generators of the Clifford algebra. The spin
generators can be decomposed w.r.t.\ $\SO{4}\times\SO{4}$ as 
\equ{
\gG_{A\,B} ~=~ \gg_{AB}\otimes \Id~,
\qquad 
\gG_{4+A\,4+B} ~=~ \Id \otimes \gg_{AB}~, 
\qquad
\gG_{A\,4+B} ~=~ \gg_A\gg \otimes \gg_B~.
}
We denote the positive chirality spin generators for both $\SO{4}$
factors as $\gG_{A\,B}^+ = \gg_{AB}^+\otimes \Id$ and 
$\gG_{4+A\,4+B}^+ = \Id \otimes \gg_{AB}^+$, respectively. 
The generators of spin $\SO{8}$ that commute with the sum
$\gG^+_{A\,B}+\gG_{4+A\,4+B}^+$ read  
\equ{
\gG_{A\,B}^- = \gg_{AB}^-\otimes \Id~,
\qquad 
\gG_{4+A\,4+B}^- = \Id \otimes \gg_{AB}^-~,
\qquad 
\gG_{A\,4+A} ~=~ \gg_A\gg \otimes \gg_A~.
}
Together these elements generate $\Sp{4}$.

\section{Flatness analysis of $\boldsymbol{\Cplx^2/\Intr_3}$ orbifold models}
\label{sc:flatness} 
\setcounter{equation}{0}

Even though it is not the most general case, we assume internal
alignment of the VEV's in hypermultiplets throughout the following
analysis. Since our purpose is to find for each of the bundle models
a realization as an heterotic orbifold theory with certain fields
taking non--vanishing VEV's, this is sufficient for our purposes. 
This analysis has been divided into U(1), SU($N$), SO($M$) and product
group flatness investigations below, as these are the gauge groups
that appear in the model listed in Table~\ref{tb:HetC2Z3}.

\subsection{U(1) flatness}

To achieve $\U{1}$ flatness we need at least two hypermultiplets. If
the hypermultiplets are charged under non--Abelian gauge multiplets
one often needs more hypermultiplets to achieve the flatness for the
other gauge symmetries as well. In particular, when one of the hyper
multiplets, is a singlet w.r.t.\ to any non--Abelian gauge symmetry,
the internal alignment phase and VEV can be adjusted to cancel the U(1) 
F--term. This means that if there are singlet hypermultiplets in the 
spectrum, U(1) flatness can always be
achieved. Hence, from Table~\ref{tb:HetC2Z3} we infer that in models 
3a, 3c and 3e U(1) flatness can always be obtained, because they
contain charged singlets. In all cases we enforce U(1) flatness only at
the end because, it just gives a single extra condition which in most
cases can be fulfilled easily by using singlets or by choosing 
relations between VEV's appropriately.

\subsection{SU(N) flatness}

The F-- and D--terms of an SU($N$) gauge group can be represented as
traceless N$\times$N matrices. It is often more convenient to not
enforce the tracelessness from the very beginning, but rather consider
the U($N$) F-- and D--terms represented by generic $N\times N$ matrices. 
Requiring that they are proportional to the identity, then enforces
SU($N$) flatness. In particular, after internal alignment has been
used, SU($N$) flatness requires that  
\equ{
\bF ~=~ \bar f\, \Id_N~,
}
where $\bar f$ is some complex number. For a single hypermultiplet 
$\gf = (\gf_1,\gf_2)$ in the fundamental representation, in which the
$\gf_1$ is a SU($N$) fundamental and the generators take the form  
$(T^{mn})_{jk}=\delta^m_{j}\delta^n_{k}$, the relation cannot be
satisfied. Indeed, employing matrix notation we have 
\equ{
\bF ~=~ \ga_\gf\, \gf_1 \bgf_1~.
}
This has determinant zero, and trace equal to $\bgf_1\gf_1$, but then
the above requirement implies that $\gf_1$ vanishes identical. Notice
that an additional charged singlet cannot help to fulfill the
flatness condition.

\subsubsection*{\centering two fundamentals}

From these considerations we conclude that at least two hypermultiplets 
$\gf$ and $\gps$ in the fundamental representation are
needed for SU($N$) flatness. Assuming internal  alignment the F--term
becomes  
\equ{
\bF ~=~ \ga_\gf\, \gf_1\bgf_1 \,+\, \ga_\gps\, \gps_1 \bgps_1~.
}
In such a case cancellation can be ensured, by choosing 
$\ga_\gf = -\ga_\gps = 1$ and $\gps_1 = \gf_1$. Hence, we conclude
that the fundamentals are aligned, and SU($N$) is broken to SU($N$-1).
Notice that all the line bundle models are realized by either having
two SU($N$) vector or U(1) charged singlet representations a
non--vanishing VEV, see the bottom part of Table~\ref{classnonab2}.

\subsubsection*{\centering one antisymmetric tensor}

For a hypermultiplet $A=(A_1,A_2)$ in the antisymmetric
representation of SU($N$), i.e.\ $A^{mn} = - A^{nm}$, the flatness
condition can be written as
\equ{ 
\bF ~=~ \ga_A\, A_1 A_1^\dag~. 
}
Using a SU($N$-1) transformations we can bring the matrix $A_1$ to a
standard form with only entries around the diagonal
\equ{
A_1 ~=~ \pmtrx{a_1 \, \ge &  &  \\ & a_2 \, \ge \\  & & \ddots}~,
\qquad 
\ge ~=~ \pmtrx{0 & 1 \\ \sm 1 & 0}~. 
}
The F--term matrix $\bF$ is then a diagonal matrix. Given this, when N
is odd, the last row and column of $A_1$ are all zero, hence the 
F--flatness implies that $A_1$ is zero entirely. When $N=2n$ even, 
the absolute values $|a_i|$ of the eigenvalues of $A_1$ are all equal; the
corresponding gauge symmetry breaking is 
SU($2n$)$\rightarrow$Sp($2n$). 
As can be see in Table~\ref{classnonab2} this possibility has been used
to obtain the bundle model with instanton number $(0,0,0,4)$ from the
heterotic orbifold 3c, given in Table~\ref{tb:HetC2Z3}.

\subsubsection*{\centering one antisymmetric tensor and one fundamental}

Next consider the situation with one antisymmetric tensor
$A=(A_1,A_2)$  and a fundamental $\gf=(\gf_1,\gf_2)$ of SU($N$). For $N$ is
even we find a previous case back in which only the antisymmetric
tensor has a VEV. For $N=2n+1$ a new possibility arises because the
fundamental $\gf_1$ can precisely be non--vanishing in the direction
where the anti--symmetric matrix in the skew--diagonal form is totally
vanishing: 
\equ{
\gf_1 ~=~ \pmtrx{c \\ 0 \\ \vdots}~, 
\qquad 
A_1 ~=~ \pmtrx{0 &  &  \\ & a_1 \, \ge \\  & & \ddots}~. 
}
The F--flatness then requires that the phases $\ga_\gf=\ga_A$ and all
entries have equal absolute values: $|c|=|a_i|$. The corresponding
symmetry breaking is $\SU{2n+1} \ra \Sp{2n}$. The bundle model with
instanton number $(0,0,0,2)$ and line bundle vector $(2,0^{15})$, see
Table~\ref{classnonab2}, can be realized in this way from the
heterotic orbifold model 3b of Table~\ref{tb:HetC2Z3}.

\subsubsection*{\centering two antisymmetric tensors (and a fundamental)}

When two antisymmetric tensors $A=(A_1,A_2)$ and $B=(B_1,B_2)$ take
non--vanishing VEVs the F--term reads
\equ{
\bF ~=~ \ga_A\, A_1 A_1^\dag \,+\, \ga_B\, B_1 B_1^\dag ~. 
}
Using an SU($N$-1) transformation we can only bring one into the form where all
entries except those immediately off the diagonal vanish. Only when
both $A_1$ and $B_1$ are skew--diagonal, the off--diagonal entries of
$\bF$ all vanish. We see that for N is even there are two classes of
solutions: 
\equ{
\arry{l l}{
\ga_A ~=~ \ga_B ~=~ 1: & |a_i|^2 \,+\, |b_i|^2 ~=~ r^2~, \\[1ex] 
\ga_A ~=~ \sm \ga_B ~=~ 1: & |a_i|^2 \,-\, |b_i|^2 ~=~ r^2~. \\[1ex] 
}
}
When $N$ is odd only the second solution is available for $r = 0$.

Depending on whether some of the eigenvalues are equal and non--zero,
zero or different, the gauge symmetry breaking varies. These different
possibilities are continously connected in the moduli space because
they are obtained from varying some of these eigenvalues. 
For the $\SU{8}$ gauge group of heterotic model
3c, see Table~\ref{tb:HetC2Z3}, the possible unbroken gauge groups
range among $\Sp{8}, \Sp{6}\times \SU{2}, \Sp{4}\times\SU{4}$ and 
$\SU{2}\times\SU{6}$ are realized as bundle models, see 
Table~\ref{classnonab2}. Again the model with the largest unbroken gauge
group, $\Sp{8}$, i.e.\ the model with instanton numbers $(0,0,0,4)$,
has another realization using only twisted states. In this case there
are many other unbroken gauge groups possible that do not occur in
Table~\ref{classnonab2} as we discuss in the main text.

The twisted and untwisted anti--symmetric tensors in
heterotic models 3b and 3c have different U(1) charges, see
Table~\ref{tb:HetC2Z3}. This means that alone they cannot achieve both
SU(N) and U(1)--flat configurations; an extra charged field is
needed. In model 3c there exists a twisted charged singlet.  For model
3b we can use one of the two $(\rep{1},\rep{5})$ to find the unbroken
gauge groups  $\Sp{4}$ and $\SU{2}^2$. The moduli space of both
SU(5) and U(1)--flat configurations thus combines the results of this
and the previous paragraph: 
\equ{
|b_1|^2 ~=~ |c|^2 \,-\, |a_1|^2~,
\qquad 
|b_2|^2 ~=~ |c|^2 \,-\, |a_2|^2~,
\qquad 
 |c|^2  ~=~\frac 52\, \big( |a_1|^2 \,+\, |a_2|^2 \big)~. 
}
Therefore generically the surviving gauge group is $\SU{2}^2$, however
when $|a_1|=|a_2|$ the symmetry is enhanced to $\Sp{4}$. This
corresponds to the  bundle model with  instanton numbers 
$(0,0,0,2)$ and line bundle vector $(2,0^{15})$, see
Table~\ref{classnonab2}. 
The generic situation describes the model with instanton number
$(0,0,0,1)$ and line bundle vector $(2,1^2,0^{13})$.

\subsection{SU($N$)$\times$SU(2)$\times$SU(2)$'$--flatness}

The heterotic model 3e of Table~\ref{tb:HetC2Z3}
has gauge group SU(14)$\times$SU(2)$\times$SU(2)$'$. Apart from the two
SU(2) doublets, the twisted spectrum contains a
$(\rep{14},\rep{2},\rep{1})$. Using similar arguments as presented for
a single fundamental of SU($N$) one concludes that a VEV for this state alone
is impossible. Therefore, combined
SU(14)$\times$SU(2)$\times$SU(2)$'$--flat configurations 
are only possible, if we give the untwisted $(\rep{14},\rep{2},\rep{2})$,
and the twisted $(\rep{14},\rep{2},\rep{1})$ and
$(\rep{1},\rep{1},\rep{2})$ VEV's simultaneously. Denoting the SU(14),
SU(2) and SU(2)$'$ indices as $a=1,\ldots N$, $i=1,2$ and $\ga=1,2$,
respectively, these hypermultiplets are $\gf_{a i \ga}$, $\gps_{a i}$
and $\gch_\ga$. The F--terms read: 
\equ{
\bF_N ~=~ \ga\, \gf_{ai\ga} \bgf_{bi\ga} +
\gb\, \gps_{ai} \bgps_{bi}~, 
~~
\bF_2 ~=~ \ga\, \gf_{ai\ga} \bgf_{aj\ga} + 
\gb\, \gps_{ai} \bgps_{aj}~, 
~~ 
\bF_2' ~=~ \ga\, \gf_{ai\ga} \bgf_{ai\gb} +  
\gg\, \gch_{\ga} \bgch_{\gb}~,  
}
with $\ga, \gb$ and $\gg$ the alignment phases. Let $v_a$ be an
arbitrary non--vanishing SU($N$) fundamental, and let $e_i = \gd_{i1}$ and
$\tilde e_{i} = \gd_{i2}$ be the standard basis vectors in two
dimensions. When we take $\ga=\gg=-\gb$, we can find two flat
solutions. The first one has 
\equ{
\gf_{ai\ga} ~=~ v_a \, e_i\, e_\ga~, 
\qquad 
\gps_{ai} ~=~ v_a \,e_i~,
\qquad 
\gch_\ga ~=~ c\, \tilde e_\ga~,
}
with $|c|^2 = |v|^2$ and surviving gauge group SU($N$-1). This is the
blowup realization of the bundle model with instanton numbers
$(0,0,0,2)$ and line bundle vector 
$\frac 12(\overbrace{1^2,\sm 1^2}\,,1^{12})$ of 
Table~\ref{classnonab2}. The other solution involves a second SU($N$)
fundamental $w_a$ which is independent of the first, say $w\cdot v =0$, so
that the configuration  
\equ{
\gf_{ai\ga} ~=~ \big(v_a \, e_i \,+\, w_a\, \tilde e_i \big)\, e_\ga~, 
\qquad 
\gps_{ai} ~=~ v_a \,e_i \,+\, w_a\, \tilde e_i ~,
\qquad 
\gch_\ga ~=~ c\, \tilde e_\ga~,
}
with $|c|^2 = |v|^2+|w|^2$ can be constructed, the unbroken gauge
group is then SU($N$-2). This leads to the second bundle model with a
spinorial line bundle vector (i.e.\ with instanton number $(0,0,0,1)$
and line bundle vector $\frac 12(\overbrace{1,\sm 1}\,,1^{12},3^2)$).

\subsection{SO($M$)$\times$SU($N$)--flatness}

In the analysis so far we only considered VEV's for representations of
SU($N$) groups. As Table~\ref{tb:HetC2Z3} models also includes SO($M$), we
have to analyze SO($M$)--flatness issues as well. The group SO($M$) has
antisymmetric generators $T^{mn} = - T^{nm}$, with $m,n=1,\ldots
M$. We see from this table that we only need the vector and spinor
representations of SO($M$) groups.

In the vector representation of SO($M$) the generators take the form: 
$(T^{mn})_{ij}=\delta^m_i \delta^n_j-\delta^m_j \delta^n_i$, so that
for a contraction with two vectors $\gf$ and $\gps$ we have 
\equ{
(\phi\wedge\psi)^{mn} ~=~ 
\phi^i T^{mn}_{ij} \psi^j ~=~ \gf^m \gps^n - \gf^n\gps^m~. 
}
Hence, we can efficiently use the index--free formalism of 2-forms
with the wedge product $\wedge$. As a warm up, we first consider
flatness for a single hypermultiplet in the vector representation,
containing the two complex vectors $\phi^i_1$ and $\phi^i_2$. We have
the following $\bF$ and $D$--term two--forms 
\equ{
\bF ~=~ \phi_1\wedge \phi_2~,
\qquad
D ~=~ \phi_1\wedge \bar \phi_1-\bar \phi_2\wedge \phi_2~. 
}
As usual we assume internal alignment so that the D--term vanish
automatically and the F--term becomes 
\equ{
\bF ~=~ \ga_\gf\, \gf_1 \wedge \bgf_1 
~=~ \ga_\gf\, Re\wedge Im ~=~ 0~, 
}
where we split $\phi_1$ into its the real and imaginary parts 
$\phi_1=Re+i Im$. Hence, F--flatness is satisfied when $Re = \pm Im$,
$Re=0$ or $Im=0$, but in any case the gauge group is broken to
SO($M$-1).

According to Table~\ref{tb:HetC2Z3} model 3b has a SO(22) vector, so
SO--flatness can be achieved in the way just described. But this
twisted state also carries U(1) charge and hence at least another
charged field needs to take a non--vanishing VEV. Since all the other
states are also charged under SU(5) we find complicated VEV
configurations.

\subsection*{\centering bi--fundamental and vectors of both groups}

The first configuration of this type corresponds to the bundle
model with instanton number $(0,0,0,1)$ and line bundle 
vector $(\overbrace{1,\sm 1}\,,1^4,0^{10})$  given in~\ref{classnonab2}: Its
non--vanishing fields are bi--fundamental $\gf=(\gf_1,\gf_2)$, 
the SO--vector $\gps=(\gps_1,\gps_2)$ and the SU--fundamental
$\gch=(\gch_1,\gch_2)$, and their internal alignment phases are $\ga$,
$\gb$ and $\gg$, respectively. Their VEVs are assigned as 
\equ{
(\gf_1)_{ai} ~=~ v_a\, e_i~,
\qquad 
(\gps_1)_a ~=~ w_a~, 
\qquad 
(\gch_1)_i ~=~ c\, e_i~, 
}
where $v$ and $w$ are two real vectors that are perpendicular to ensure 
SO--flatness, and $c$ a complex constant. The phases are chosen as
$\ga=-\gb=-\gg=1$, then $|c| = v^2$ for SU--flatness, and $w^2$ is
adjusted to also have U(1) flatness.  If we instead take $w=v$, still all
flatness conditions can be fulfilled but we end up with the bundle
model with instanton number $(1,0,0,0)$ and the non--overlapping line
bundle vector  $(1^4,0^{12})$.

\subsection*{\centering bi--fundamental, SO--vector and SU--antisymmetric tensor}

Another configuration employs the VEV's of the bi--fundamental, one
SU--antisymmetric tensor and an SO--vector. When their VEV assignments
are given by 
\equ{
(\gf_1)_a ~=~ v_a~,
\qquad
(\gps_1)_{ai} ~=~ v_a \, e_i~, 
\qquad 
A_1 ~=~ \pmtrx{a\, \ge & & \\ & a\, \ge & \\ & & 0}~, 
}
flatness is achieved provided that $|c| = |v|$. The resulting
unbroken gauge group reads SO(21)$\times$Sp(4), hence this gives the
blowup realization of the bundle model with instanton number
$(1,0,0,2)$ that has no additional line bundle embedding.

\subsection*{\centering bi--fundamental, SO--vector and two
SU--antisymmetric tensors}

A third type of configurations combines VEV's of the
bi--fundamental, two SU--antisymmetric tensors, and the
SO--vector. Their VEV's are  
\equ{
(\gf_1)_{ai} ~=~ v_a\, e_i~,
\qquad 
(\gps_1)_a ~=~ w_a~, 
\qquad 
A_1 ~=~ \pmtrx{a\, \ge & & \\ & 0 & \\ & & 0}~, 
\qquad 
B_1 ~=~ \pmtrx{0 & & \\ & b\, \ge & \\ & & 0}~, 
}
and lead to the symmetry breaking from SO(22)$\times$SU(5) to
SO(20)$\times$SU(2)$^2$ provided that $v$ and $w$ are
perpendicular. This corresponds to the bundle 
model with instanton number $(0,0,0,2)$ and line bundle vector
$(\overbrace{1,\sm 1}\,,1^2,0^{12})$ in Table~\ref{classnonab2}. 
When the two vectors $v$ and $w$ are equal, the gauge symmetry is only
broken to SO(21)$\times$SU(2)$^2$, i.e.\ we recover the bundle model
with instanton number $(1,0,0,1)$ and non--overlapping line bundle
vector $(1^2,0^{14})$.

\subsection*{\centering two bi--fundamentals}

In the final configuration we consider, there are two
bi--fundamentals. We can view the components of the bi--fundamentals
 as SO--vectors $a, b, c$ and $d$ 
\equ{
\gs_1 ~=~ \pmtrx{a  & b}~, 
\qquad 
\gps_1 ~=~ \pmtrx{c & d}~. 
}
The SO--flatness is fulfilled when these vectors are all real.
SU--flatness gives the conditions  
\equ{
\bar a b \,\pm\, \bar c d ~=~ 0~, 
\qquad 
|a|^2 \,\pm\, |c|^2 ~=~ |b|^2 \,\pm\, |d|^2~,
}
where the $\pm$--sign distinguishes between two possible alignments. 
By taking the vectors perpendicular when they are not proportional,
one solves the first equation trivially. With this class of VEV
configurations various bundle models are obtained: When all four
vectors are perpendicular we end up with gauge group SO(24), i.e.\ the
bundle model with instanton number $(0,0,0,2)$ and fully overlapping
line bundle vector $(\overbrace{1^2,\sm 1^2}\,,0^{12})$. With three
non--vanishing vectors with $|b|^2 = |a|^2+|c|^2$ and $d=0$ we have
the gauge group $SO(25)$: the bundle model with $n=(1,0,0,1)$ and
$V=(\overbrace{1,-1}\,,0^{14})$. Finally when we align all four vectors, 
the gauge group is SO(27); the bundle model $n=(1,0,0,0)$
and $V=(2,0^{15})$ is found.

\bibliographystyle{paper}
{\small
\bibliography{paper}

\providecommand{\href}[2]{#2}\begingroup\raggedright\begin{thebibliography}{10}

\bibitem{Faraggi:1989ka}
A.~E. Faraggi, D.~V. Nanopoulos, and K.-j. Yuan ``A standard like model in the
  4d free fermionic string formulation'' {\em Nucl. Phys.} {\bf B335} (1990)
347.
%%CITATION = NUPHA,B335,347;%%.

\bibitem{Blumenhagen:2001te}
R.~Blumenhagen, B.~Kors, D.~Lust, and T.~Ott ``{The Standard Model from stable
  intersecting brane world orbifolds}'' {\em Nucl. Phys.} {\bf B616} (2001)
  3--33
\href{http://www.arXiv.org/abs/hep-th/0107138}{[{\tt hep-th/0107138}]}.
%%CITATION = HEP-TH/0107138;%%.

\bibitem{Cvetic:2001nr}
M.~Cvetic, G.~Shiu, and A.~M. Uranga ``{Chiral four-dimensional N = 1
  supersymmetric type IIA orientifolds from intersecting D6-branes}'' {\em
  Nucl. Phys.} {\bf B615} (2001) 3--32
\href{http://www.arXiv.org/abs/hep-th/0107166}{[{\tt hep-th/0107166}]}.
%%CITATION = HEP-TH/0107166;%%.

\bibitem{Dijkstra:2004cc}
T.~P.~T. Dijkstra, L.~R. Huiszoon, and A.~N. Schellekens ``{Supersymmetric
  Standard Model spectra from RCFT orientifolds}'' {\em Nucl. Phys.} {\bf B710}
  (2005) 3--57
\href{http://www.arXiv.org/abs/hep-th/0411129}{[{\tt hep-th/0411129}]}.
%%CITATION = HEP-TH/0411129;%%.

\bibitem{Dijkstra:2004ym}
T.~P.~T. Dijkstra, L.~R. Huiszoon, and A.~N. Schellekens ``{Chiral
  supersymmetric Standard Model spectra from orientifolds of Gepner models}''
  {\em Phys. Lett.} {\bf B609} (2005) 408--417
\href{http://www.arXiv.org/abs/hep-th/0403196}{[{\tt hep-th/0403196}]}.
%%CITATION = HEP-TH/0403196;%%.

\bibitem{Dixon:1985jw}
L.~Dixon, J.~A. Harvey, C.~Vafa, and E.~Witten ``Strings on orbifolds'' {\em
  Nucl. Phys.} {\bf B261} (1985)
678--686.
%%CITATION = NUPHA,B261,678;%%.

\bibitem{Dixon:1986jc}
L.~J. Dixon, J.~A. Harvey, C.~Vafa, and E.~Witten ``Strings on orbifolds. 2''
  {\em Nucl. Phys.} {\bf B274} (1986)
285--314.
%%CITATION = NUPHA,B274,285;%%.

\bibitem{Ibanez:1987sn}
L.~E. Ibanez, H.~P. Nilles, and F.~Quevedo ``Orbifolds and {W}ilson lines''
  {\em Phys. Lett.} {\bf B187} (1987)
25--32.
%%CITATION = PHLTA,B187,25;%%.

\bibitem{Buchmuller:2005jr}
W.~Buchmuller, K.~Hamaguchi, O.~Lebedev, and M.~Ratz ``{Supersymmetric Standard
  Model from the heterotic string}'' {\em Phys. Rev. Lett.} {\bf 96} (2006)
  121602
\href{http://www.arXiv.org/abs/hep-ph/0511035}{[{\tt hep-ph/0511035}]}.
%%CITATION = HEP-PH/0511035;%%.

\bibitem{Buchmuller:2006ik}
W.~Buchmuller, K.~Hamaguchi, O.~Lebedev, and M.~Ratz ``{Supersymmetric Standard
  Model from the heterotic string. II}'' {\em Nucl. Phys.} {\bf B785} (2007)
  149--209
\href{http://www.arXiv.org/abs/hep-th/0606187}{[{\tt hep-th/0606187}]}.
%%CITATION = HEP-TH/0606187;%%.

\bibitem{Lebedev:2006kn}
O.~Lebedev {\em et al.} ``A mini-landscape of exact {MSSM} spectra in heterotic
  orbifolds'' {\em Phys. Lett.} {\bf B645} (2007) 88--94
\href{http://www.arXiv.org/abs/hep-th/0611095}{[{\tt hep-th/0611095}]}.
%%CITATION = HEP-TH/0611095;%%.

\bibitem{Altarelli:2001qj}
G.~Altarelli and F.~Feruglio ``{SU(5) grand unification in extra dimensions and
  proton decay}'' {\em Phys. Lett.} {\bf B511} (2001) 257--264
\href{http://www.arXiv.org/abs/hep-ph/0102301}{[{\tt hep-ph/0102301}]}.
%%CITATION = HEP-PH/0102301;%%.

\bibitem{Kobayashi:2004ya}
T.~Kobayashi, S.~Raby, and R.-J. Zhang ``Searching for realistic 4d string
  models with a {P}ati-{S}alam symmetry: Orbifold grand unified theories from
  heterotic string compactification on a {Z(6)} orbifold'' {\em Nucl. Phys.}
  {\bf B704} (2005) 3--55
\href{http://www.arXiv.org/abs/hep-ph/0409098}{[{\tt hep-ph/0409098}]}.
%%CITATION = HEP-PH 0409098;%%.

\bibitem{Forste:2004ie}
S.~Forste, H.~P. Nilles, P.~K.~S. Vaudrevange, and A.~Wingerter ``Heterotic
  brane world'' {\em Phys. Rev.} {\bf D70} (2004) 106008
\href{http://www.arXiv.org/abs/hep-th/0406208}{[{\tt hep-th/0406208}]}.
%%CITATION = HEP-TH 0406208;%%.

\bibitem{Hebecker:2004ce}
A.~Hebecker and M.~Trapletti ``Gauge unification in highly anisotropic string
  compactifications'' {\em Nucl. Phys.} {\bf B713} (2005) 173--203
\href{http://www.arXiv.org/abs/hep-th/0411131}{[{\tt hep-th/0411131}]}.
%%CITATION = HEP-TH 0411131;%%.

\bibitem{Kim:2006hw}
J.~E. Kim and B.~Kyae ``Flipped {SU(5)} from {Z(12-I)} orbifold with {W}ilson
  line'' {\em Nucl. Phys.} {\bf B770} (2007) 47--82
\href{http://www.arXiv.org/abs/hep-th/0608086}{[{\tt hep-th/0608086}]}.
%%CITATION = HEP-TH/0608086;%%.

\bibitem{Donaldson:1985}
S.~Donalson ``Anti-self-dual {Y}ang--{M}ills connections over complex algebraic
  surfaces and stable vector bundles'' {\em Proc. Londan Math. Soc.} {\bf 50}
  (1985) 1--26.

\bibitem{Uhlenbeck:1986}
K.~Uhlenbeck and S.~Yau ``On the existence of {H}ermitian-{Y}ang-{M}ills
  connections in stable vector bundles'' {\em Comm. Pure and Appl. Math.} {\bf
  19} (1986) 257--293.

\bibitem{Braun:2005ux}
V.~Braun, Y.-H. He, B.~A. Ovrut, and T.~Pantev ``A heterotic standard model''
  {\em Phys. Lett.} {\bf B618} (2005) 252--258
\href{http://www.arXiv.org/abs/hep-th/0501070}{[{\tt hep-th/0501070}]}.
%%CITATION = HEP-TH 0501070;%%.

\bibitem{Braun:2005bw}
V.~Braun, Y.-H. He, B.~A. Ovrut, and T.~Pantev ``A standard model from the
  {E(8) x E(8)} heterotic superstring'' {\em JHEP} {\bf 06} (2005) 039
\href{http://www.arXiv.org/abs/hep-th/0502155}{[{\tt hep-th/0502155}]}.
%%CITATION = HEP-TH 0502155;%%.

\bibitem{Blumenhagen:2005zg}
R.~Blumenhagen, G.~Honecker, and T.~Weigand ``{Non-Abelian brane worlds: The
  heterotic string story}'' {\em JHEP} {\bf 10} (2005) 086
\href{http://www.arXiv.org/abs/hep-th/0510049}{[{\tt hep-th/0510049}]}.
%%CITATION = HEP-TH/0510049;%%.

\bibitem{Bouchard:2005ag}
V.~Bouchard and R.~Donagi ``An {SU(5)} heterotic standard model'' {\em Phys.
  Lett.} {\bf B633} (2006) 783--791
\href{http://www.arXiv.org/abs/hep-th/0512149}{[{\tt hep-th/0512149}]}.
%%CITATION = HEP-TH 0512149;%%.

\bibitem{Blumenhagen:2006ux}
R.~Blumenhagen, S.~Moster, and T.~Weigand ``{Heterotic GUT and Standard Model
  vacua from simply connected Calabi-Yau manifolds}'' {\em Nucl. Phys.} {\bf
  B751} (2006) 186--221
\href{http://www.arXiv.org/abs/hep-th/0603015}{[{\tt hep-th/0603015}]}.
%%CITATION = HEP-TH/0603015;%%.

\bibitem{Honecker:2006qz}
G.~Honecker and M.~Trapletti ``Merging heterotic orbifolds and {K3}
  compactifications with line bundles'' {\em JHEP} {\bf 01} (2007) 051
\href{http://www.arXiv.org/abs/hep-th/0612030}{[{\tt hep-th/0612030}]}.
%%CITATION = HEP-TH 0612030;%%.

\bibitem{Nibbelink:2007rd}
S.~Groot~Nibbelink, M.~Trapletti, and M.~Walter ``Resolutions of {C**n/Z(n)}
  orbifolds, their {U(1)} bundles, and applications to string model building''
  {\em JHEP} {\bf 03} (2007) 035
\href{http://www.arXiv.org/abs/hep-th/0701227}{[{\tt hep-th/0701227}]}.
%%CITATION = HEP-TH/0701227;%%.

\bibitem{Ganor:2002ae}
O.~J. Ganor and J.~Sonnenschein ``On the strong coupling dynamics of heterotic
  string theory on {C**3/Z(3)}'' {\em JHEP} {\bf 05} (2002) 018
\href{http://www.arXiv.org/abs/hep-th/0202206}{[{\tt hep-th/0202206}]}.
%%CITATION = HEP-TH/0202206;%%.

\bibitem{GrootNibbelink:2007ew}
S.~Groot~Nibbelink, H.~P. Nilles, and M.~Trapletti ``{Multiple anomalous U(1)s
  in heterotic blow-ups}'' {\em Phys. Lett.} {\bf B652} (2007) 124--127
\href{http://www.arXiv.org/abs/hep-th/0703211}{[{\tt hep-th/0703211}]}.
%%CITATION = HEP-TH/0703211;%%.

\bibitem{Nibbelink:2008tv}
S.~Groot~Nibbelink, D.~Klevers, F.~Ploger, M.~Trapletti, and P.~K.~S.
  Vaudrevange ``{Compact heterotic orbifolds in blow-up}'' {\em JHEP} {\bf 04}
  (2008) 060
\href{http://www.arXiv.org/abs/0802.2809}{[{\tt 0802.2809}]}.
%%CITATION = 0802.2809;%%.

\bibitem{Erler:1992ki}
J.~Erler and A.~Klemm ``Comment on the generation number in orbifold
  compactifications'' {\em Commun. Math. Phys.} {\bf 153} (1993) 579--604
\href{http://www.arXiv.org/abs/hep-th/9207111}{[{\tt hep-th/9207111}]}.
%%CITATION = HEP-TH/9207111;%%.

\bibitem{Aspinwall:1994ev}
P.~S. Aspinwall ``{Resolution of orbifold singularities in string theory}''
\href{http://www.arXiv.org/abs/hep-th/9403123}{[{\tt hep-th/9403123}]}.
%%CITATION = HEP-TH/9403123;%%.

\bibitem{Lust:2006zh}
D.~Lust, S.~Reffert, E.~Scheidegger, and S.~Stieberger ``Resolved toroidal
  orbifolds and their orientifolds''
\href{http://www.arXiv.org/abs/hep-th/0609014}{[{\tt hep-th/0609014}]}.
%%CITATION = HEP-TH/0609014;%%.

\bibitem{Nibbelink:2007pn}
S.~Groot~Nibbelink, T.-W. Ha, and M.~Trapletti ``{Toric Resolutions of
  Heterotic Orbifolds}'' {\em Phys. Rev.} {\bf D77} (2008) 026002
\href{http://www.arXiv.org/abs/0707.1597}{[{\tt 0707.1597}]}.
%%CITATION = 0707.1597;%%.

\bibitem{Eguchi:1978xp}
T.~Eguchi and A.~J. Hanson ``{Asymptotically Flat Selfdual Solutions to
  Euclidean Gravity}'' {\em Phys. Lett.} {\bf B74} (1978)
249.
%%CITATION = PHLTA,B74,249;%%.

\bibitem{Gibbons:1979zt}
G.~W. Gibbons and S.~W. Hawking ``{Gravitational Multi - Instantons}'' {\em
  Phys. Lett.} {\bf B78} (1978)
430.
%%CITATION = PHLTA,B78,430;%%.

\bibitem{Eguchi:1980jx}
T.~Eguchi, P.~B. Gilkey, and A.~J. Hanson ``{Gravitation, Gauge Theories and
  Differential Geometry}'' {\em Phys. Rept.} {\bf 66} (1980)
213.
%%CITATION = PRPLC,66,213;%%.

\bibitem{Douglas:1996sw}
M.~R. Douglas and G.~W. Moore ``{D-branes, Quivers, and ALE Instantons}''
\href{http://www.arXiv.org/abs/hep-th/9603167}{[{\tt hep-th/9603167}]}.
%%CITATION = HEP-TH/9603167;%%.

\bibitem{Bianchi:1996zj}
M.~Bianchi, F.~Fucito, G.~Rossi, and M.~Martellini ``{Explicit Construction of
  Yang-Mills Instantons on ALE Spaces}'' {\em Nucl. Phys.} {\bf B473} (1996)
  367--404
\href{http://www.arXiv.org/abs/hep-th/9601162}{[{\tt hep-th/9601162}]}.
%%CITATION = HEP-TH/9601162;%%.

\bibitem{ConradThesis}
J.~Conrad {\em Orbifolds and Kaluza-Klein-Monopoles in Heterotic E8xE8 String
  Theory Preserving Eight Supercharges}.
\newblock PhD thesis Bonn University 2001.
\newblock http://thp.uni-bonn.de/nilles/db/thesis/conrad.pdf.

\bibitem{Eguchi:1978gw}
T.~Eguchi and A.~J. Hanson ``{Selfdual Solutions to Euclidean Gravity}'' {\em
  Ann. Phys.} {\bf 120} (1979)
82.
%%CITATION = APNYA,120,82;%%.

\bibitem{Sen:1997js}
A.~Sen ``{Dynamics of multiple Kaluza-Klein monopoles in M and string theory}''
  {\em Adv. Theor. Math. Phys.} {\bf 1} (1998) 115--126
\href{http://www.arXiv.org/abs/hep-th/9707042}{[{\tt hep-th/9707042}]}.
%%CITATION = HEP-TH/9707042;%%.

\bibitem{Ruback:1986ag}
P.~J. Ruback ``The motion of {K}aluza-{K}lein monopoles'' {\em Commun. Math.
  Phys.} {\bf 107} (1986)
93--102.
%%CITATION = CMPHA,107,93;%%.

\bibitem{'tHooft:1974qc}
G.~'t~Hooft ``Magnetic monopoles in unified gauge theories'' {\em Nucl. Phys.}
  {\bf B79} (1974)
276--284.
%%CITATION = NUPHA,B79,276;%%.

\bibitem{Nakahara:1990th}
M.~Nakahara ``Geometry, topology and physics''. Bristol, UK: Hilger (1990) 505
  p. (Graduate student series in physics).

\bibitem{Atiyah:1978ri}
M.~F. Atiyah, N.~J. Hitchin, V.~G. Drinfeld, and Y.~I. Manin ``{Construction of
  instantons}'' {\em Phys. Lett.} {\bf A65} (1978)
185--187.
%%CITATION = PHLTA,A65,185;%%.

\bibitem{Dorey:2002ik}
N.~Dorey, T.~J. Hollowood, V.~V. Khoze, and M.~P. Mattis ``{The calculus of
  many instantons}'' {\em Phys. Rept.} {\bf 371} (2002) 231--459
\href{http://www.arXiv.org/abs/hep-th/0206063}{[{\tt hep-th/0206063}]}.
%%CITATION = HEP-TH/0206063;%%.

\bibitem{Kronheimer:1989zs}
P.~B. Kronheimer ``{The Construction of ALE spaces as hyperKahler quotients}''
  {\em J. Diff. Geom.} {\bf 29} (1989)
665--683.
%%CITATION = JDGEA,29,665;%%.

\bibitem{Kronheimer:1990}
P.~B. Kronheimer and H.~Nakajima ``{Yang-Mills instantons on ALE gravitational
  instantons}'' {\em Math. Ann.} {\bf 288} (1990)
263--307.
%%CITATION = JDGEA,29,665;%%.

\bibitem{Nakajima:1990}
P.~B. Kronheimer and H.~Nakajima ``{Moduli spaces of anti-self-dual connections
  on ALE gravitational instantons}'' {\em Invent. Math.} {\bf 102} (1990)
267--303.
%%CITATION = JDGEA,29,665;%%.

\bibitem{Fucito:2004ry}
F.~Fucito, J.~F. Morales, and R.~Poghossian ``{Multi instanton calculus on ALE
  spaces}'' {\em Nucl. Phys.} {\bf B703} (2004) 518--536
\href{http://www.arXiv.org/abs/hep-th/0406243}{[{\tt hep-th/0406243}]}.
%%CITATION = HEP-TH/0406243;%%.

\bibitem{Wilczek:1976uy}
F.~Wilczek ``{Inequivalent embeddings of SU(2) and instanton interactions}''
  {\em Phys. Lett.} {\bf B65} (1976)
160--162.
%%CITATION = PHLTA,B65,160;%%.

\bibitem{Vandoren:2008xg}
S.~Vandoren and P.~van Nieuwenhuizen ``{Lectures on instantons}''
\href{http://www.arXiv.org/abs/0802.1862}{[{\tt 0802.1862}]}.
%%CITATION = 0802.1862;%%.

\bibitem{Fulton}
W.~Fulton {\em Introduction to Toric Varieties}.
\newblock Princeton University Press 1993.

\bibitem{Oda}
T.~Oda {\em Convex Bodies and Algebraic Geometry: An Introduction to the Theory
  of Toric Varieties}.
\newblock Springer 1988.

\bibitem{Bouchard:2007ik}
V.~Bouchard ``Lectures on complex geometry, {C}alabi-{Y}au manifolds and toric
  geometry''
\href{http://www.arXiv.org/abs/hep-th/0702063}{[{\tt hep-th/0702063}]}.
%%CITATION = HEP-TH/0702063;%%.

\bibitem{Aldazabal:1997wi}
G.~Aldazabal, A.~Font, L.~E. Ibanez, A.~M. Uranga, and G.~Violero
  ``Non-perturbative heterotic {D = 6,4, N = 1} orbifold vacua'' {\em Nucl.
  Phys.} {\bf B519} (1998) 239--281
\href{http://www.arXiv.org/abs/hep-th/9706158}{[{\tt hep-th/9706158}]}.
%%CITATION = HEP-TH 9706158;%%.

\bibitem{Ibanez:1997td}
L.~E. Ibanez and A.~M. Uranga ``D = 6, {N} = 1 string vacua and duality''
\href{http://www.arXiv.org/abs/hep-th/9707075}{[{\tt hep-th/9707075}]}.
%%CITATION = HEP-TH 9707075;%%.

\bibitem{Witten:1985xe}
E.~Witten ``Global gravitational anomalies'' {\em Commun. Math. Phys.} {\bf
  100} (1985)
197.
%%CITATION = CMPHA,100,197;%%.

\bibitem{Freed:1986zx}
D.~S. Freed ``Determinants, torsion, and strings'' {\em Commun. Math. Phys.}
  {\bf 107} (1986)
483--513.
%%CITATION = CMPHA,107,483;%%.

\bibitem{Honecker:2006dt}
G.~Honecker ``{Massive U(1)s and heterotic five-branes on K3}'' {\em Nucl.
  Phys.} {\bf B748} (2006) 126--148
\href{http://www.arXiv.org/abs/hep-th/0602101}{[{\tt hep-th/0602101}]}.
%%CITATION = HEP-TH/0602101;%%.

\bibitem{Slansky:1981yr}
R.~Slansky ``Group theory for unified model building'' {\em Phys. Rept.} {\bf
  79} (1981)
1--128.
%%CITATION = PRPLC,79,1;%%.

\bibitem{Green:1984bx}
M.~B. Green, J.~H. Schwarz, and P.~C. West ``Anomaly free chiral theories in
  six-dimensions'' {\em Nucl. Phys.} {\bf B254} (1985)
327--348.
%%CITATION = NUPHA,B254,327;%%.

\bibitem{Erler:1993zy}
J.~Erler ``Anomaly cancellation in six-dimensions'' {\em J. Math. Phys.} {\bf
  35} (1994) 1819--1833
\href{http://www.arXiv.org/abs/hep-th/9304104}{[{\tt hep-th/9304104}]}.
%%CITATION = HEP-TH 9304104;%%.

\bibitem{Douglas:1997de}
M.~R. Douglas, B.~R. Greene, and D.~R. Morrison ``{Orbifold resolution by
  D-branes}'' {\em Nucl. Phys.} {\bf B506} (1997) 84--106
\href{http://www.arXiv.org/abs/hep-th/9704151}{[{\tt hep-th/9704151}]}.
%%CITATION = HEP-TH/9704151;%%.

\bibitem{Seiberg:1996vs}
N.~Seiberg and E.~Witten ``{Comments on String Dynamics in Six Dimensions}''
  {\em Nucl. Phys.} {\bf B471} (1996) 121--134
\href{http://www.arXiv.org/abs/hep-th/9603003}{[{\tt hep-th/9603003}]}.
%%CITATION = HEP-TH/9603003;%%.

\bibitem{Duff:1996rs}
M.~J. Duff, R.~Minasian, and E.~Witten ``{Evidence for Heterotic/Heterotic
  Duality}'' {\em Nucl. Phys.} {\bf B465} (1996) 413--438
\href{http://www.arXiv.org/abs/hep-th/9601036}{[{\tt hep-th/9601036}]}.
%%CITATION = HEP-TH/9601036;%%.

\bibitem{Witten:1984dg}
E.~Witten ``Some properties of {O}(32) superstrings'' {\em Phys. Lett.} {\bf
  B149} (1984)
351--356.
%%CITATION = PHLTA,B149,351;%%.

\bibitem{Georgi:1982jb}
H.~Georgi ``{L}ie algebras in particle physics. from isospin to unified
  theories'' {\em Front. Phys.} {\bf 54} (1982)
1--255.
%%CITATION = FRPHA,54,1;%%.

\end{thebibliography}\endgroup
}
\end{document}